\documentclass[pdflatex,sn-basic]{sn-jnl}

\usepackage{natbib}
\usepackage{graphicx}%
\usepackage{multirow}%
\usepackage{amsmath,amssymb,amsfonts}%
\usepackage{amsthm}%
\usepackage{mathrsfs}%
\usepackage[title]{appendix}%
\usepackage{xcolor}%
\usepackage{textcomp}%
\usepackage{manyfoot}%
\usepackage{booktabs}%
\usepackage{algorithm}%
\usepackage{algorithmicx}%
\usepackage{algpseudocode}%
\usepackage{listings}%
\usepackage[nonumberlist,nosuper]{glossaries} 
\setacronymstyle{long-short} 

\usepackage{xcolor, cancel, soul}
\usepackage[normalem]{ulem}
\definecolor{g}{RGB}{0,160,0}
\setstcolor{g}
\definecolor{b}{RGB}{0,0,245}
\setstcolor{b}

\theoremstyle{thmstyleone}%
\theoremstyle{thmstyletwo}%

\theoremstyle{thmstylethree}%

\newacronym{mo}{M-O}{magneto-optical} 
\newacronym{nlte}{non-LTE}{non-local thermodynamical equilibrium} 
\newacronym{lte}{LTE}{local thermodynamical equilibrium} 
\newacronym{prd}{PRD}{partial frequency redistribution} 
\newacronym{crd}{CRD}{complete frequency redistribution} 
\newacronym{1d}{1D}{one dimensional} 
\newacronym{3d}{3D}{three dimensional} 
\newacronym{rt}{RT}{radiative transfer} 
\newacronym{los}{LOS}{line-of-sight} 
\newacronym{wfa}{WFA}{weak-field approximation} 
\newacronym{fwhm}{FWHM}{full width at half-maximum} 
\newacronym{tic}{TIC}{Tenerife Inversion Code} 
\newacronym{uv}{UV}{ultraviolet} 
\newacronym{se}{SE}{statistical equilibrium} 
\newacronym{svd}{SVD}{singular value decomposition} 
\newacronym{psf}{PSF}{point spread function} 

\newcommand{\fref}[1]{Fig.\,\ref{#1}} 
\newcommand{\eref}[1]{Eq.\,(\ref{#1})} 
\newcommand{\tref}[1]{Table\,\ref{#1}} 
\newcommand{\sref}[1]{Section\,\ref{#1}}

\begin{document} 

\title[Article Title]{Inferring Solar Magnetic fields From the 
Polarization of the Mg {\sc II} h and k Lines} 


\author*[1]{\fnm{Hao} \sur{Li}}\email{lihao01@nssc.ac.cn} 

\author[2,3]{\fnm{Tanaus\'u} \sur{del Pino Alem\'an}}\email{tanausu@iac.es} 

\author[2,3,4]{\fnm{Javier} \sur{Trujillo Bueno}}\email{jtb@iac.es} 

\affil[1]{\orgdiv{State Key Laboratory of Solar Activity and Space Weather}, 
\orgname{National Space Science Center, Chinese Academy of Sciences}, 
\orgaddress{\street{}\postcode{100190} \city{Beijing}, 
\state{}\country{People's Republic of China}}} 

\affil[2]{\orgdiv{}\orgname{Instituto de Astrof\'{\i}sica de Canarias}, 
\orgaddress{\street{}\postcode{E-38205} \city{La Laguna}, 
\state{Tenerife}, \country{Spain}}} 

\affil[3]{\orgdiv{Departamento de Astrof\'\i sica}, 
\orgname{Universidad de La Laguna}, 
\orgaddress{\street{}\postcode{E-38206} \city{La Laguna}, 
\state{Tenerife}, \country{Spain}}} 

\affil[4]{\orgdiv{Consejo Superior de Investigaciones Cient\'{\i}ficas}, 
\orgaddress{\country{Spain}}} 



\abstract{The polarization of the Mg {\sc II} h and k lines holds significant 
diagnostic potential for measuring chromospheric magnetic fields, which 
are crucial for understanding the physical processes governing the energy 
transport and dissipation in the solar upper atmosphere, as well as the subsequent 
heating of the chromosphere and corona. 
The Chromospheric Layer Spectropolarimeter was launched twice in 2019 and 
2021, successfully acquiring spectropolarimetric observations across 
the Mg {\sc II} h and k lines. 
The analysis of these observations confirms the capability of these lines for 
inferring magnetic fields in the upper chromosphere. 
In this review, we briefly introduce the physical mechanisms behind the 
polarization of the Mg {\sc II} h and k lines, including the joint action of 
the Zeeman and Hanle effects, the magneto-optical effect, partial frequency redistribution,
and atomic level polarization. 
We also provide an overview of recent progress in the interpretation of the Stokes 
profiles of the Mg {\sc II} h and k lines.} 


\maketitle

\section{Introduction}\label{sec1} 

The solar chromosphere is a critical interface layer, coupling the 
relatively cool photosphere with the much hotter corona \citep{Carlsson2019ARA&A}. 
Despite being significantly cooler than the corona, due to its higher density, 
the chromosphere requires a comparatively larger amount of energy input
to maintain its temperature of approximately 10~000 K \citep{Withbroe1977ARA&A}. 
Moreover, there is evidence suggesting that the energy responsible for heating 
the corona to a temperature exceeding 1~MK may originate in the upper 
chromosphere and transition region, rather than in the corona 
itself \citep{Aschwanden2007ApJ}.

Several mechanisms have been proposed to explain the coronal heating problem, 
such as magnetic reconnection, and dissipation of Alfv{\'e}nic waves 
\citep{Walsh2003A&ARv,VanDoorsselaere2020SSRv}, which are all closely tied to magnetic fields. 
In addition, the plasma $\beta$ (the ratio of gas to magnetic pressure) decreases 
with height as the density falls in the chromosphere \citep{Gary2001SoPh}. 
As a result, the magnetic field plays a dominant role in shaping the dynamics and 
structures of the plasma in the upper chromosphere, where the plasma $\beta$ is low. 
Therefore, determining the strength and geometry of the magnetic field
in the solar chromosphere is essential for understanding how energy is 
transported and dissipated in the chromosphere and transition region, 
which ultimately results in the chromospheric and coronal heating.

Except for radio techniques (\citealt{Casini2017SSRv}; \citealt{Chen2020NatAs}; 
\citealt{Chen2025ApJ}), coronal seismology methods 
\citep{Yang2020Sci,Yang2020ScChE,Yang2024Sci}, and diagnostics based on 
magnetic field-induced transitions  
\citep{Li2015ApJL,Landi2020ApJ,Chen2021ApJ}, 
the inversion of the polarized solar spectrum remains one of our primary means 
to infer the magnetic field in the solar atmosphere 
\citep[see the reviews by][]{delToroIniesta2016LRSP,Lagg2017SSRv,delaCruz2017SSRv}. 
The polarization of the electromagnetic radiation emerging 
from the solar atmosphere encodes information about the physical 
properties of the emitting plasma, including the magnetic field. 
Thus, the interpretation of the polarized solar spectrum is essential 
for revealing the characteristics of the magnetic field, which in turn 
enhances our understanding of the physical processes taking place 
in the solar atmosphere. 
To date, routine measurements of the photospheric magnetic fields have been 
achieved by exploiting the spectral line polarization 
produced by the Zeeman effect \citep{Zeeman1896VMKAN}. 
However, determining magnetic fields in the chromosphere (encoded, especially
in the upper chromosphere, in strong \gls*{uv} lines) remains a significant 
challenge, particularly outside of active regions, 
where the linear polarization induced by the Zeeman effect is often too 
weak to be detected due to the small Zeeman splitting compared to the 
relatively large line widths
\citep[see the review by][]{TrujilloBueno2022ARA&A}. 

In the solar atmosphere, the anisotropic incident 
radiation can lead to population imbalances among the magnetic 
sublevels of atomic levels, giving rise to linearly polarized spectral lines. 
This linear polarization, dubbed scattering polarization,
is subsequently modified in the presence of a magnetic field 
through the Hanle effect \citep{Hanle1924ZPhy,TrujilloBueno2001ASPC}, 
which offers significant potential for detecting weak magnetic fields 
\citep[see the monographs by][]{Stenflo1994ASSL,LL04}. 
Among the spectral lines originating in the upper solar chromosphere, 
strong \gls*{uv} resonance lines such as the H {\sc I} Ly-$\rm \alpha$
\citep{TrujilloBueno2011ApJL,TrujilloBueno2012ApJL,Belluzzi2012bApJL,
Stepan2012ApJL,Ishikawa2014ApJ,Stepan2015ApJ} 
and the Mg {\sc II} h and k lines 
\citep{Belluzzi2012ApJL,AlsinaBallester2016ApJL,delPinoAleman2016ApJL,
delPinoAleman2020ApJ,Hofmann2025ApJ}  
are considered particularly promising for magnetic field diagnostics 
\citep{TrujilloBueno2017SSRv}. 

Motivated by these theoretical investigations on the polarization induced 
by scattering processes, and the Hanle and Zeeman effects, 
a series of sounding rocket experiments were carried out to observe the 
polarization of these strong \gls*{uv} lines. 
These experiments are the Chromospheric 
Lyman-Alpha SpectroPolarimeter 
\citep[CLASP,][]{Kobayashi2012ASPC} 
and the Chromospheric LAyer SpectroPolarimeter 
\citep[CLASP2, ][]{Narukage2016SPIE,Song2018SPIE}, launched in 2015 and 
2019, respectively, as well as the reflight of CLASP2 (CLASP2.1) in 2021. 
These missions successfully recorded the intensity and linear polarization 
of the H {\sc I} Ly-$\rm \alpha$ line \citep{Kano2017ApJ,TrujilloBueno2018ApJL}, 
as well as the full Stokes vector of the Mg {\sc II} h and k lines 
\citep{Ishikawa2021SciA,Rachmeler2022ApJ}. 
The subsequent analysis of these data has demonstrated the diagnostic 
potential of these and other near \gls*{uv} lines for probing chromospheric 
magnetic fields, particularly the Mg {\sc II} h and k lines and the 
nearby \gls*{uv} lines, of which the 
circular polarization induced by the Zeeman effect is also observable in 
active regions with enough signal-to-noise ratio
\citep{Ishikawa2021SciA,Ishikawa2025ApJ,AfonsoDelgado2023aApJ,AfonsoDelgado2025ApJ,
Li2023ApJ,Li2024bApJ,Li2024aApJ,Song2025ApJ}. 

This review focuses on the polarization of the Mg {\sc II} 
h and k lines. 
In \sref{sec2}, we briefly introduce the main aspects regarding the formation of 
these lines. 
In \sref{sec3}, we provide an overview of the key physical mechanisms 
impacting their polarization, including the joint action of 
the Zeeman and Hanle effects, \gls*{mo} effects, \gls*{prd} effects, and atomic 
polarization. 
In \sref{sec4}, we present the forward modeling of the polarization of these 
lines in \gls*{nlte} conditions.  
Non-LTE inversions and the \gls*{wfa} for inferring the 
magnetic field from the Stokes profiles are introduced 
in \sref{sec5} and \sref{sec6}, respectively. 
In \sref{sec7}, we highlight recent progress in the interpretation of
the spectropolarimetric observations of the Mg {\sc II} 
h and k lines obtained by CLASP2 and CLASP2.1. 
Finally, a summary and future perspectives are provided in \sref{sec8}.

\section{Formation of the Mg {\sc II} h and k lines}\label{sec2} 

\begin{figure*}[htp]
  \center
  \includegraphics[width=1.\textwidth]{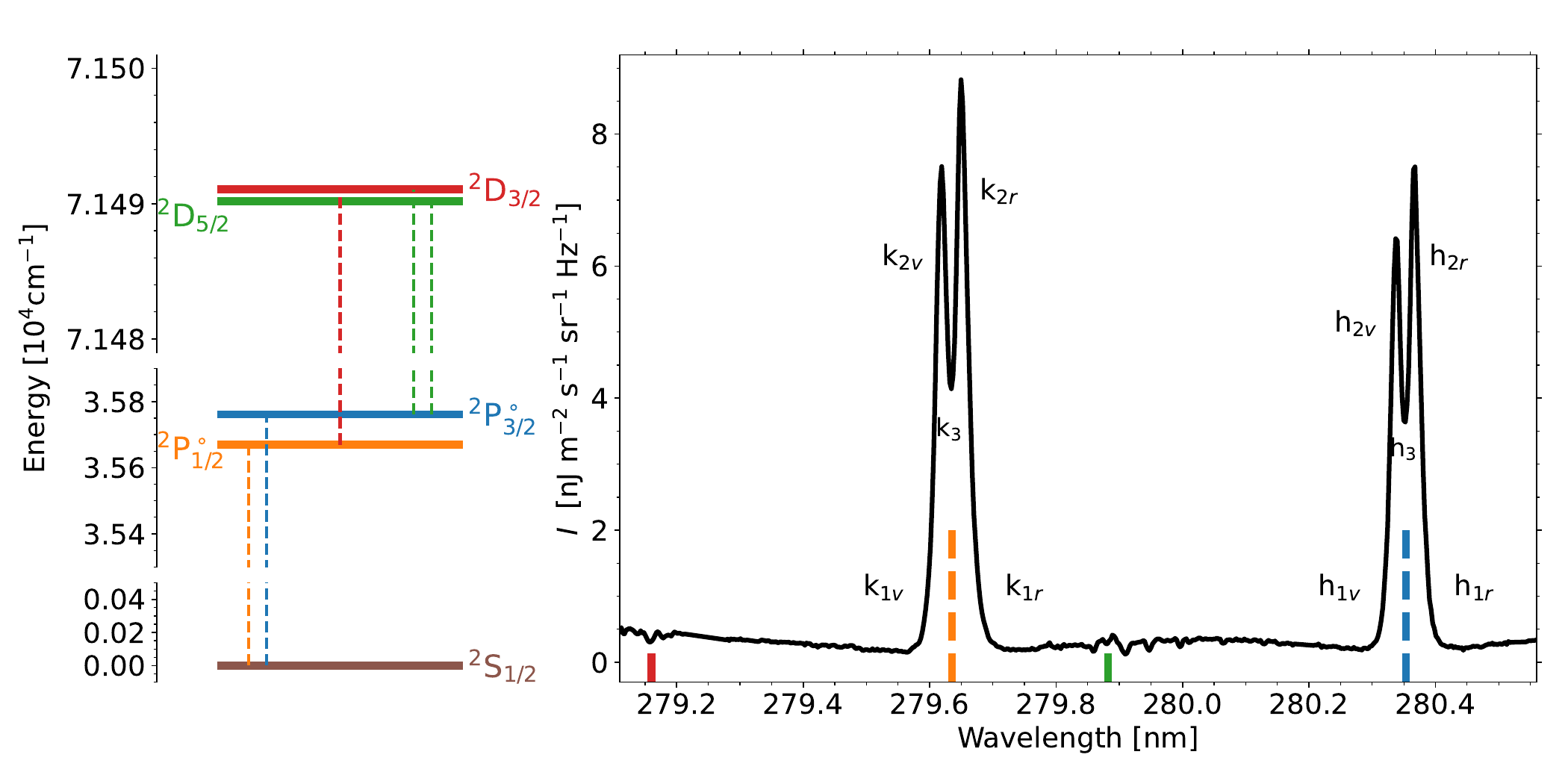}
  \caption{Left panel: Grotrian diagram of the Mg {\sc II} atomic model, 
  illustrating the transitions corresponding the Mg {\sc II} h and k 
  lines at 280.353~nm and 279.635~nm, respectively, as well as the 
  subordinate lines at 279.160~nm, 279.875~nm, and 279.882~nm. 
  Right panel: Intensity profiles of the Mg {\sc II} h and k lines, 
  as well as the subordinate lines observed in an active region by the IRIS satellite. 
  The vertical dashed curves indicate the center of these lines. 
  The subscripts 3, 2, and 1 denote the line cores, 
  emission peaks, and near wings of the h and k lines, respectively.} 
  \label{fig1}
\end{figure*}

Due to the relatively low ionization potential of the Mg {\sc I} atom, the 
majority of the magnesium atoms in the photosphere and chromosphere exists 
in its singly ionized state, 
making Mg {\sc II} the dominant ion in these 
layers \citep{Carlsson2012A&A}. 
Furthermore, due to its relatively large elemental abundance in the solar 
atmosphere \citep{Asplund2009ARA&A}, the Mg {\sc II} h and k resonance lines 
at 280.353~nm and 279.635~nm, respectively, are among the strongest lines in the solar 
spectrum \citep{Judge2022ApJ}. 
The Grotrian diagram in the left panel of \fref{fig1} illustrates the 
energy levels relevant for the Mg {\sc II} h and k lines, as well as 
the subordinate lines. 
The corresponding wavelengths and the Einstein coefficients for spontaneous 
emission are listed in \tref{tab1}. 
The h and k lines share a common lower level, the ground state of 
Mg {\sc II}, while their upper levels, which are also the lower levels of 
the subordinate lines, belong to the same $\rm ^2P^\circ$ term 
and are separated by a relatively small energy gap. 

The Mg {\sc II} h and k lines cannot be observed using 
ground-based instruments, due to the ozone absorption in the \gls*{uv} 
wavelength range. 
Since the 1950s, these lines have been observed via rocket experiments 
\citep[e.g.,][]{Johnson1953ApJ} and from space telescopes.  
The right panel of \fref{fig1} presents the intensity profile of the Mg {\sc II} 
h, k, and subordinate lines acquired by the Interface Region Imaging Spectrograph 
\citep[IRIS,][]{DePontieu2014SoPh}. 

The wings of the Mg {\sc II} h and k lines form in the photosphere 
and lower chromosphere, while the two emission peaks form in the middle chromosphere. 
The minima outside these peaks, dubbed h$_1$ (h$_{1v}$ and h$_{1r}$) and 
k$_1$ (k$_{1v}$ and k$_{1r}$) for the h and k lines, respectively, form close to 
the temperature minimum in standard semi-empirical models \citep{Vernazza1981ApJS}. 
The emission peaks are referred to as h$_{2v}$ and h$_{2r}$ for the h line, 
and k$_{2v}$ and k$_{2r}$ for the k line, respectively. 
As seen in the right panel of \fref{fig1}, these lines typically exhibit 
absorption features in the 
line cores due to absorption in the upper chromosphere \citep[e.g.,][]{Schmit2015ApJ}. 
However, in plage regions the absorption can disappear. 
The line centers are dubbed h$_3$ and k$_3$ for the h and k lines, 
respectively. 

\begin{table}[htp]
  \caption{Transition wavelength ($\lambda$), Einstein coefficient for spontaneous emission
  ($A_{ul}$), and effective Land{\'e} factors ($\bar{g}$) of the Mg {\sc II} 
  h, k, and subordinate lines, as well as the critical magnetic field strength 
  for the lines that are sensitive to the Hanle effect.} 
  \label{tab1}
  \begin{tabular}{c c c c c c c c}
  \hline\hline
  Transition                                 & $\lambda$ (vacuum)     & $A_{ul}$          & $\bar{g}$ & $B_{\rm H}$ \\
                                             & (nm)                   & ($\rm s^{-1}$)    &           & (G)         \\
  \hline
  $\rm ^2P^\circ_{1/2}\rightarrow ^2S_{1/2}$ & 280.353 (h line)     & $2.57\times 10^8$ & 1.33      & -           \\
  $\rm ^2P^\circ_{3/2}\rightarrow ^2S_{1/2}$ & 279.635 (k line)     & $2.60\times 10^8$ & 1.17      & 22          \\
  $\rm ^2D_{3/2}\rightarrow ^2P^\circ_{1/2}$ & 279.160                 & $4.01\times 10^8$ & 0.83      & 57          \\
  $\rm ^2D_{3/2}\rightarrow ^2P^\circ_{3/2}$ & 279.875 (blended) & $7.98\times 10^7$ & 1.07      & 11          \\
  $\rm ^2D_{5/2}\rightarrow ^2P^\circ_{3/2}$ & 279.882 (blended) & $4.79\times 10^8$ & 1.10      & 45          \\
  \hline
  \end{tabular}
  \begin{tablenotes}
  \item \textbf{Note.} The Einstein emission coefficients are taken from the
  NIST database \citep{NIST}. 
  The effective Land{\'e} factors are computed by assuming LS coupling. 
  \end{tablenotes}
\end{table}

Due to the slightly larger Einstein coefficient for absorption, $B_{lu}$, 
of the k line with respect to that of the h line, 
the height where the optical depth is unity for the emission peaks of 
the k line is slightly higher than for the h line. 
The source function of the k line at the height where the optical depth is unity 
is also larger than that of the h line \citep{Leenaarts2013aApJ}. 
Consequently, the k line shows comparatively larger intensities at its peaks
\citep{Linsky1970PASP}, as seen in the right panel of \fref{fig1}. 

The subordinate Mg {\sc II} lines at 279.160~nm, 279.875~nm, and 279.882~nm 
lie in the wings of the h and k lines, and are much weaker.
Two of these subordinate lines (279.875~nm, and 279.882~nm) are blended. 
Typically, these lines appear in absorption, but can exhibit emission 
if there is a local heating in the lower chromosphere, 
where they are formed \citep{Pereira2015ApJ}. 
The formation regions of the Mg {\sc II} h, k, and subordinate lines can be 
analyzed in detail using their response functions 
\citep[RFs; ][]{LandiInnocenti1977A&A} with 
respect to the model parameters \citep[e.g., ][]{delaCruz2016ApJL}. 
Note that their formation heights vary slightly between different atmosphere 
models. 
Overall, the Mg {\sc II} h, k, and subordinate lines are sensitive to the 
plasma properties across a wide range of heights throughout the solar 
chromosphere, making them excellent probes for the
thermodynamic \citep{Uitenbroek1997SoPh,Leenaarts2013bApJ,Pereira2013ApJ} 
and magnetic \citep{Belluzzi2012ApJL,delPinoAleman2020ApJ} structures in the chromosphere, particularly in the upper layers 
just below the transition region.

\section{Polarization physics of the Mg {\sc II} 
h and k lines}\label{sec3}

Due to the relatively low plasma density, the \gls*{lte} approximation 
is generally not valid in the solar chromosphere. 
The atomic populations do not follow the 
Saha--Boltzmann equations. Therefore, the effects of \gls*{nlte} need to be 
taken into account when modeling the Mg {\sc II} h and k lines. 
In order to infer the magnetic field from their polarization,
it is essential to fully understand the 
physical processes that give rise to the polarization.
In this section, we briefly introduce such mechanisms, including the 
Zeeman and Hanle effects, atomic level polarization, quantum interference 
between $J$-states, and partial frequency redistribution.
All these effects must be accounted for when modeling the polarization 
of the Mg {\sc II} h and k lines 
\citep{Belluzzi2012ApJL,Belluzzi2014A&A,AlsinaBallester2016ApJL,delPinoAleman2016ApJL,delPinoAleman2020ApJ}.

\subsection{Zeeman effect}\label{sec3.1}

Through the Zeeman effect, more than one hundred years 
ago \citet{Hale1908ApJ} discovered the presence of 
magnetic fields in sunspots. 
Today, the Zeeman effect is routinely employed to infer
magnetic fields in the solar photosphere \citep[e.g., the reviews by][]{
delToroIniesta2016LRSP,Lagg2017SSRv,delaCruz2017SSRv,Kleint2017SSRv}. 
The Zeeman effect arises from the interaction of a magnetic field
with the magnetic moment of the electrons in the atom, leading to the 
energy splitting of the 2$J$+1 sublevels of any
given atomic level with total angular momentum $J$. 
These sublevels are characterized by the magnetic quantum number, 
$M=-J,-J+1,\dots,J$.
In the so-called Zeeman regime, this energy splitting is linearly
proportional to the magnetic field strength.

\begin{figure*}[htp]
  \center
  \includegraphics[width=1.\textwidth]{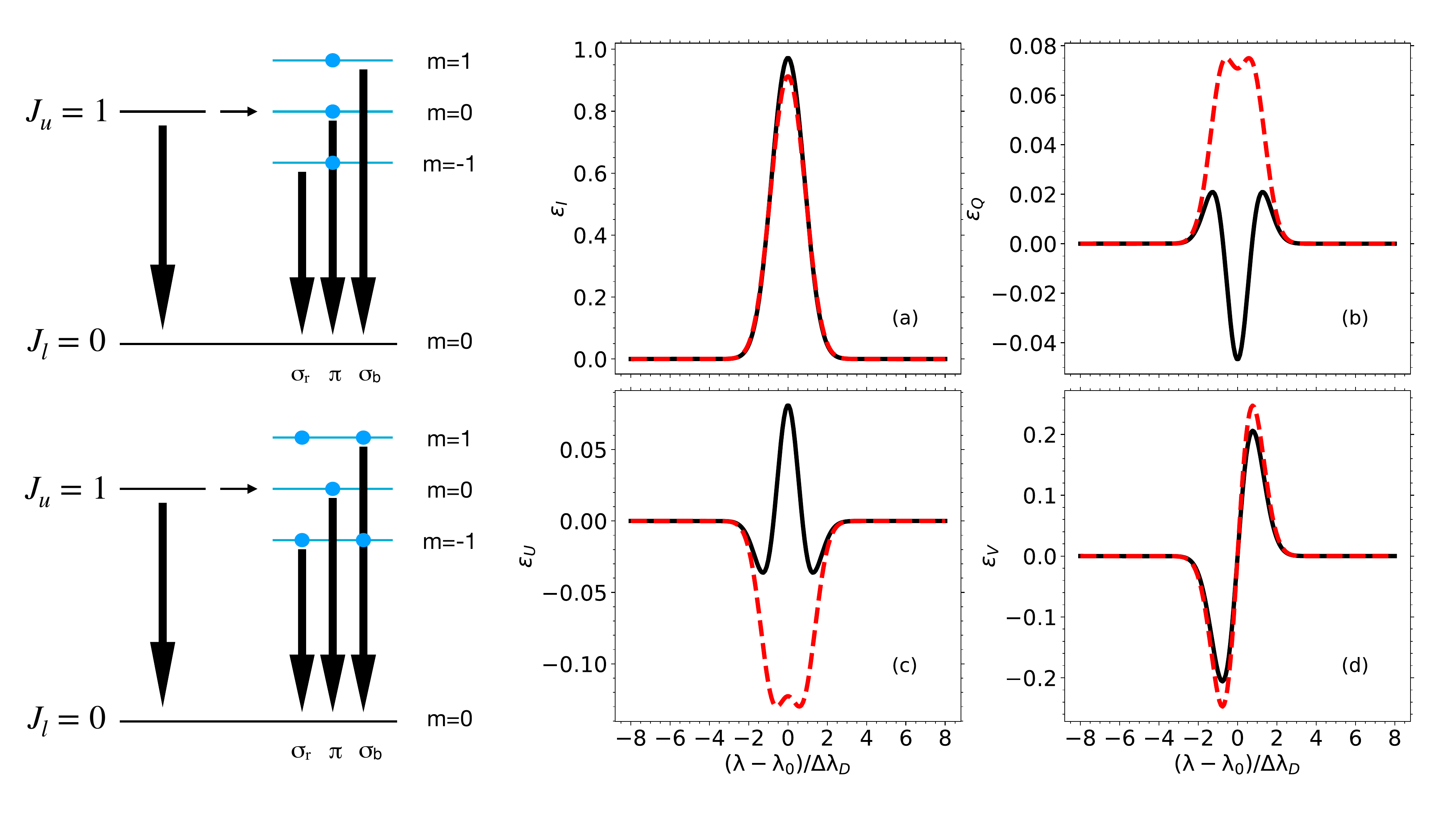}
  \caption{The two panels to the left illustrate the Zeeman splitting 
  of the energy levels corresponding to a transition between atomic levels 
  with $J_u=1$ and $J_l=0$. 
  The top left panel shows the case of an evenly distributed population among the 
  magnetic sublevels, while the bottom left panel shows an uneven distribution. 
  The relative populations are indicated by the blue circles on the blue horizontal 
  lines representing the magnetic sublevels. 
  The quantization $\mathcal{Z}$--axis is taken along the magnetic field.
  The central and right columns show the calculated emission profiles for
  the Stokes $I$, $Q$, $U$, and $V$ parameters, respectively, assuming a Zeeman 
  splitting of $\Delta\lambda_D/2$. 
  The reference direction for positive Stokes $Q$ is defined by an 
  azimuth $\chi$ with respect to the transverse component of the magnetic field. 
  When $\chi= 0$, i.e. when the reference direction is parallel to the transverse component of 
  the magnetic field, the Stokes U signal vanishes. 
  The black solid and red dotted curves correspond to the emissivity with evenly 
  and unevenly
  distributed populations, respectively. 
  The Stokes $I$ profile in the evenly distributed population case is normalized 
  to its integral over wavelength. 
  In the unevenly distributed population case, the relative intensities of the 
  $\pi$ and $\sigma$ components are adjusted according to the population ratios shown 
  in the bottom left panel (i.e., the ratio of populations for the
  the magnetic sublevels M = -1, 0, and 1 is 2:1:2).}
  \label{fig2}
\end{figure*}

The polarization that the Zeeman effect introduces in a spectral line is due 
to the wavelength shift of the transitions between the $M$ magnetic sublevels. 
In the Zeeman regime, this shift is given by \citep{LL04}, 
\begin{equation}\label{eq1}
  \Delta\lambda = \Delta\lambda_B(g_\ell M_\ell-g_uM_u) 
  = 4.6686\times10^{-13}\lambda_0^2B(g_\ell M_\ell-g_uM_u), 
\end{equation}
where $\Delta\lambda$ is wavelength shift in {\AA}, $\lambda_0$ is the central 
wavelength in {\AA} of the unsplit transition, $B$ is magnetic field 
strength in gauss,
$g_u$ and $g_\ell$ are the Land{\'e} factors of the upper and lower atomic
levels, respectively, and
$M_u$ and $M_\ell$ are the magnetic quantum numbers of the upper and lower 
magnetic sublevels, respectively. 
For dipole-type transitions, only those transitions with $\Delta M=M_u-M_\ell=0,\pm 1$ 
are permitted. 
The transition with $\Delta M = 0$ is referred to as the $\pi$ component, 
and those with $\Delta M = \pm 1$ are the $\sigma_b$ and $\sigma_r$ 
components, respectively. 
The magnetic splitting of a given level is often characterized by the equation 
for a classical Zeeman triplet \citep[e.g., ][Sect. 3.3]{LL04},  
\begin{equation}\label{eq2} 
  \Delta\lambda = \bar{g}\Delta\lambda_B,
\end{equation}
where $\bar{g}$ is the effective Land{\'e} factor.

The top left panel of \fref{fig2} illustrates the Zeeman splitting of the 
energy levels corresponding to a transition from an upper level with $J_u=1$ 
to a lower level with $J_l=0$. 
The upper level splits into three magnetic sublevels, whereas the lower 
level remains unsplit because it has $J_l=0$. 
The resulting spectral line consists of $\pi$ and $\sigma$ components, 
each with a different polarization state.
In the absence of a magnetic field, these components coincide in 
wavelength, and their polarization states cancel out, resulting in no 
net polarization (assuming an even population distribution among the 
magnetic sublevels). 
However, when a magnetic field is present, the energy splitting causes each 
component to have a different wavelength, preventing their polarization states 
from cancelling out and producing a net observable polarization.

The observed polarization depends on the relative
direction between the magnetic field and the \gls*{los}.
For instance, the Zeeman induced polarization is purely circular 
when the \gls*{los} is parallel to the magnetic field, 
and purely linear when it is perpendicular. 
For any other viewing direction, the observed polarization is elliptical. 

The emission coefficients for the Stokes parameters of an atom with Zeeman splitting
are given by \citep[e.g.,][]{delToroIniesta2003}, 
\begin{subequations}\label{eq3}
  \begin{align}
  \epsilon_I & = C[\phi_p\sin^2\theta+\frac{\phi_b+\phi_r}{2}(1+\cos^2\theta)], \label{eq3a} \\
  \epsilon_Q & = C(\phi_p-\frac{\phi_b+\phi_r}{2})\sin^2\theta\cos 2\chi, \label{eq3b} \\
  \epsilon_U & = C(\phi_p-\frac{\phi_b+\phi_r}{2})\sin^2\theta\sin 2\chi, \label{eq3c} \\
  \epsilon_V & = C(\phi_r-\phi_b)\cos\theta, \label{eq3d}
  \end{align}
\end{subequations}
where $C$ is a constant which depends on the plasma properties. 
$\phi_{p,b,r}$ correspond to the emission profiles of 
the $\pi$, $\sigma_b$, and $\sigma_r$ components, respectively. 
$\theta$ is the inclination of the magnetic field with respect to 
the \gls*{los}, and  
$\chi$ is the azimuth of the transverse component of the magnetic field 
with respect to the positive Stokes $Q$ reference direction.

The black solid curves in the middle and right panels of \fref{fig2} show the 
emission coefficients calculated corresponding to $\theta=\pi/3$ and $\chi=\pi/3$.
Gaussian profiles with a Doppler width $\Delta\lambda_D$ were assumed for 
$\phi_p$, $\phi_r$, and $\phi_b$. 
The wavelength shifts of the $\sigma$ components were assumed to be
$\Delta\lambda_D/2$. 
Additionally, the population was assumed to be evenly distributed 
among the magnetic sublevels.

Although in the chosen example the transverse magnetic field component is stronger 
than the longitudinal component, the resulting linear polarization is weaker 
than the circular polarization. 
This is because, in the weak field regime ($\Delta\lambda_B\ll\Delta\lambda_D$), 
the amplitude of the circular polarization induced by the Zeeman effect scales 
linearly with the ratio of the Zeeman splitting to the Doppler width, whereas the 
linear polarization scales with the square of this ratio \citep{LL04}. 
Even if this argument is strictly correct only in the weak-field regime,
the relation remains essentially valid for stronger fields. 
In the chromosphere, where the magnetic fields are typically weaker than in the 
photosphere and spectral lines tend to be wider, this ratio usually is much smaller 
than unity, making the detection of transverse magnetic 
fields especially challenging.

\subsection{Atomic level polarization}\label{sec3.2}

The black solid curves in \fref{fig2} were computed under the assumption that 
the atomic level populations are evenly distributed among magnetic sublevels. 
However, this assumption is not always suitable, especially in rarefied
plasmas \citep{TrujilloBueno2001ASPC}.
The population imbalance, illustrated in the bottom left panel of 
\fref{fig2}, modifies the relative intensities of the $\pi$ and $\sigma$ components, 
leading to a noticeable difference in the Stokes profiles of the emergent 
spectral line radiation.
Even in the absence of a magnetic field, such population imbalance is capable of
producing polarization, dubbed scattering polarization.
In the solar atmosphere, the population imbalances arise from anisotropic
optical pumping \citep{Cohen-Tannoudji1966PrOpt,Happer1972RvMP},  
a process in which the absorption and scattering of anisotropic radiation leads to 
unequal populations among the magnetic sublevels.  
Furthermore, the population imbalances can occur among the magnetic sublevels of 
both the upper and
lower levels \citep{TrujilloBueno1997ApJ,TrujilloBueno1999ASSL,TrujilloBueno2002Natur,MansoSainz2003PhRvL}.
The magnetic sublevels can be not only unevenly populated, but
they can also be quantum mechanically coupled \citep{LL04}. The term
atomic level polarization refers to both the population imbalance and to the quantum
coherence among the magnetic sublevels of any given atomic level.

The scattering polarization signals are particularly significant near the solar limb. 
Spectropolarimetric observations throughout the solar spectrum 
from 3165~{\AA} to 9950~{\AA} at $\mu=0.1$ ($\mu=\cos\theta$, with
$\theta$ the heliocentric angle) reveal a wealth of linearly polarized 
spectral structures \citep{Stenflo1983aA&AS,Stenflo1983bA&AS,Stenflo1997A&A}, 
which was dubbed the Second Solar Spectrum by \citet{Ivanov1991ASIC}. 
An atlas of the Second Solar Spectrum, covering the 
wavelengths from the near UV to the near infrared, was subsequently obtained 
by \citet{Gandorfer2000sss,Gandorfer2002sss,Gandorfer2005sss}, 
using the Zurich Imaging Polarimeter 
\citep[ZIMPOL,][]{Povel1995OptEn,Povel2001ASPC,Gandorfer2004A&A}.

The polarization of the atomic levels can be described using the 
irreducible spherical tensor components of the density matrix, 
$\rho^K_Q(\alpha J,\alpha 'J')$ \citep{Fano1957RvMP,Omont1977PQE}, 
where $\alpha$ and $\alpha'$ 
indicate the atomic states (i.e., electronic configuration). 
In most cases, quantum interference between the sublevels pertaining 
to different atomic levels (i.e., $\alpha J\neq \alpha 'J'$) is negligible.
The coherence between $J$-states pertaining to the same atomic term will
be discussed in \sref{sec3.4}. 

For $\alpha J = \alpha' J'$, the rank $K=0,1,\dots,2J$, and $Q=-K,-K+1,\dots,K$. 
The component $\rho^0_0(\alpha J)$ is proportional to the
total population of the atomic level.
The tensor components with $K=2$ are referred to as the alignment components 
and contribute to the linear polarization, 
while those with $K=1$ are referred to as the orientation components 
and contribute to the circular polarization.
The components with $Q=0$ describe the distribution of the population among 
the magnetic sublevels, while those with $Q\neq 0$ describe the 
quantum coherence between magnetic sublevels and are generally complex numbers.

The density matrix tensor components with $K=1$ cannot usually be excited by 
unpolarized incident radiation, but they can be generated through 
alignment--orientation conversion in the presence of an electric field
\citep{Casini2005PhRvA,Rochester2012PhRvA}, making them a potential means
for the diagnostic such fields \citep{,Anan2024NatCo}.
For atomic levels with $J\leqslant1/2$, the maximum rank 
$K_{\rm max} \leqslant 1$, implying that no alignment components exist. 
Therefore, there cannot be linear scattering polarization in transitions 
between levels with $J\leqslant1/2$, which is the case for the Mg {\sc II} h line.

\subsection{Hanle effect}\label{sec3.3}

The atomic level polarization can be modified by a magnetic field, which in
turn leads to a modification of the line scattering polarization. 
This mechanism is dubbed Hanle effect after the discovery in the laboratory 
by \citeauthor{Hanle1924ZPhy} (\citeyear{Hanle1924ZPhy}; see also 
\citealt{TrujilloBueno2001ASPC}). 
This effect causes both a rotation and a change in the amplitude of the 
linear scattering polarization.
The Hanle effect arises from the modification of the quantum 
coherence between the magnetic sublevels of an atomic state.

The modification of the atomic level polarization by a magnetic 
field due to the Hanle effect for a two-level atom (neglecting 
stimulated emission and assuming an unpolarized lower level)
can be described by \citep{LL04}, 
\begin{equation}\label{eq4}
  \rho^K_Q(J_u) = \frac{1}{1+iQH_u}[\rho^K_Q(J_u)]_{B=0},
\end{equation}
where the quantization axis is taken parallel to the magnetic field vector.  
$\rho^K_Q(J_u)$ are the irreducible spherical tensor
components of the density matrix in the presence of a magnetic field, while the 
subscript $B=0$ denotes the field-free case. 
The dimensionless Hanle parameter, $H_u$, is defined as 
$H_u=0.879\times 10^7t_{\rm life}gB$, 
where $t_{\rm life}$ is the lifetime (in seconds) of the atomic level under 
consideration, $g$ is its Land{\'e} factor, and $B$ is the magnetic field 
strength (in gauss).

From \eref{eq4}, the Hanle effect does not operate when the magnetic 
field is too weak ($H_u\ll 1$), and it saturates for sufficiently strong magnetic fields 
($H_u\gg 1$), in which case the quantum coherence terms with $Q\ne 0$ vanish.
The Hanle effect is thus sensitive to magnetic field strengths 
within an approximate range of $0.2B_{\rm H} < B < 5B_{\rm H}$, where
$B_{\rm H}$ is the critical Hanle field given by \citep[e.g., ][]{TrujilloBueno2022ARA&A}, 
\begin{equation}\label{eq5}
  B_{\rm H}=\frac{1.137\times 10^{-7}}{t_{\rm life}g}. 
\end{equation}
The critical magnetic field of the Mg {\sc II} k line is listed in \tref{tab1}, 
with $t_{\rm life} $ being roughly estimated as $A_{ul}^{-1}$. 
Hence, the Hanle effect of the Mg {\sc II} k line is roughly sensitive to 
magnetic field strengths between 4~G and 110~G, approximately. 
In contrast to the Zeeman effect, the Hanle effect is sensitive to 
tangled magnetic fields at subresolution scales \citep{Stenflo1982SoPh}. 
This has enabled the detection of a substantial amount of hidden magnetic 
energy in the quiet Sun \citep{TrujilloBueno2004Natur}.

\subsection{$J$-state interference}\label{sec3.4} 

As mentioned in Section\,\ref{sec3.2}, the density matrix components
$\rho^K_Q(\alpha J,\alpha 'J')$ with $\alpha J\neq \alpha 'J'$, which describe 
the quantum interference between $J$-states, are usually neglected. 
However, within a given atomic term ($\alpha = \alpha'$), when the energy
separation between the levels $\alpha J$ 
and $\alpha J'$ is sufficiently small, these components can become significant. 
In particular, when \gls*{prd} effects are important, $J$-state interference can 
have a substantial impact on the spectral line polarization. 
This is the case for spectral lines such as the Ca {\sc II} H and K lines, 
the Mg {\sc II} h and k lines, and the Na {\sc I} $\rm D_1$ and $\rm D_2$ 
lines \citep{Auer1980A&A,Stenflo1980A&A,Belluzzi2012ApJL}. 
Generally, $J$-state quantum interference leads to polarization features
in the wings of spectral lines \citep{Belluzzi2011ApJ}. 

The impact of quantum interference between the upper levels of the Mg {\sc II} h 
and k lines in the absence of a magnetic field was first predicted 
by \cite{Auer1980A&A}, who assumed coherent scattering in the observer's frame, 
an approximation reasonable only in the far wings. 
Their modeling predicted that the Stokes $Q$ profile exhibits strong positive 
polarization signals toward the red wing of the h line and toward the blue wing of 
the k line, while becoming negative in the spectral region 
between the two lines. 
This prediction was later confirmed by \citet{Belluzzi2012ApJL}, 
considering $J$-state interference and \gls*{prd} effects, and 
by \citet{delPinoAleman2016ApJL,delPinoAleman2020ApJ} and \citet{AlsinaBallester2022A&A}, 
including arbitrary magnetic fields.

The Ultraviolet Spectrometer and Polarimeter \citep[UVSP,][]{Woodgate1980SoPh} 
aboard the Solar Maximum Mission \citep{Bohlin1980SoPh} acquired observations 
of the linear polarization across the Mg {\sc II} h and k lines. 
While the first analysis by \cite{Henze1987SoPh} did not reveal significant linear 
polarization in the spectral region of the wings between these two lines, 
a subsequent reanalysis by \citet{MansoSainz2019ApJL} detected negative Stokes $Q$ polarization. 
This finding was clearly confirmed by the high-precision spectropolarimetric 
observations acquired by the CLASP2 sounding rocket experiment \citep{Rachmeler2022ApJ}. 
These unprecedented observations also confirmed the theoretical prediction 
of \citet{Belluzzi2012ApJL} for the whole spectral range of the Mg {\sc II} 
h and k lines, including the near wings around the centers of these lines.

\subsection{Magneto-optical effects}\label{sec3.5}

The significant scattering polarization Stokes $Q$ signals in the line wings 
of the Mg {\sc II} h and k lines can give rise to significant Stokes $U$ signals through 
the \gls*{mo} effects in the presence of a longitudinal magnetic field.
The \gls*{rt} equations for Stokes $Q$ and $U$ are given by, 
\begin{subequations}\label{eq6}
\begin{align}
\frac{dQ}{ds} & = \epsilon_Q-\eta_QI-\rho_VU+\rho_UV-\eta_IQ \simeq \epsilon_Q-\rho_VU-\eta_IQ, \label{eq6a} \\
\frac{dU}{ds} & = \epsilon_U-\eta_UI+\rho_VQ-\rho_QV-\eta_IU \simeq \epsilon_U+\rho_VQ-\eta_IU, \label{eq6b}
\end{align}
\end{subequations}
where $\epsilon_{I,Q,U,V}$ and $\eta_{I,Q,U,V}$ are the emission 
and absorption coefficients for the four Stokes parameters, respectively.
The $\rho_{Q,U,V}$ terms are responsible for the \gls*{mo} effect. 
For the Mg {\sc II} h and k lines, $\rho_V$ is negligible 
at the line center, but becomes significant in the line wings, as its
spectral shape results from the superposition of antisymmetric dispersion
profiles \citep[see][]{AlsinaBallester2016ApJL}.
Consequently, the terms $\rho_VU$ and $\rho_VQ$ lead to 
a rotation of the linear polarization introducing sensitivity in 
the line wings of Stokes $Q$ and $U$ to the presence of magnetic fields 
as weak as those that activate the Hanle effect at the center of the k 
line \citep[see][]{AlsinaBallester2016ApJL}.  
The manifestation of the \gls*{mo} effects discussed here
requires a significant scattering polarization signal in the wings of the
spectral lines. Therefore, these \gls*{mo} effects are intimately associated with 
the $J$-state interference and \gls*{prd} effects, which is why we often distinguish 
them from the well-known \gls*{mo} effect caused by the 
Zeeman effect in the presence of strong magnetic fields \citep{LL04}, which is 
significant in the line core region without the need of scattering polarization.

\subsection{Partial frequency redistribution}\label{sec3.6}

A rigorous quantum physics framework for describing the Hanle effect 
was built by \citet{LandiInnocenti1972SoPh}, \citet{Bommier1978A&A}, 
\citet{Bommier1980A&A}, and \citet{LandiInnocenti1983aSoPh,
LandiInnocenti1983bSoPh,LandiInnocenti1984SoPh,LandiInnocenti1985SoPh} under 
the assumption of \gls*{crd}. 
The \gls*{crd} assumption is valid in highly collisional plasmas, where
collisions destroy the coherence between the incident and the emitted
radiations, or when this coherence is negligible because the incident 
radiation is spectrally-flat across the spectral line under consideration, also
known as the flat-spectrum approximation \citep{Casini2008pps}. In both cases
the scattering processes can be treated as a temporal succession of statistically
independent first-order processes.
\citet{Sahal-Brechot1974aA&A,Sahal-Brechot1974bA&A,Sahal-Brechot1977ApJ}, 
and \citet{Casini1999ApJ} 
extended the formalism from electric to magnetic 
dipole transitions. 
More recently, \citet{Casini2025ApJ} extended 
it to electric and magnetic quadrupolar transitions.
In this quantum mechanical theoretical framework, the \gls*{se} equations 
governing the density matrix, 
and the \gls*{rt} equations for the Stokes parameters
are derived self-consistently 
from the theory of quantum electrodynamics. 

For strong resonance lines such as the Mg {\sc II} h and k lines, 
the Ca {\sc II} H and K lines, or the H {\sc I} Ly-$\alpha$ line, 
the coherence between the absorbed and the emitted photons is 
not negligible. 
This coherence leads to a dependence of the emission profile on 
both the incoming and outcoming frequencies of the photon 
\citep{Hummer1962MNRAS,Mihalas1978}, 
which is referred to as \gls*{prd}. 
The effects of \gls*{prd} have been taken into account in forward modeling 
and inversion codes in \gls*{1d} plane parallel atmospheric models, such 
as RH \citep{Uitenbroek2001ApJ,Pereira2015A&A},  
STiC \citep{delaCruz2016ApJL,delaCruz2019A&A},
and HanleRT-TIC \citep{delPinoAleman2016ApJL,Li2022ApJ}, 
with the latter accounting for the joint action of the Zeeman and Hanle effects, 
and scattering polarization. 
\gls*{prd} effects have also been accounted for in 
\gls*{3d} \gls*{nlte} \gls*{rt} calculations by \citet{Sukhorukov2017A&A},
albeit without polarization. 
Accounting simultaneously for \gls*{prd} effects, atomic level
polarization, the Hanle and Zeeman effects, in \gls*{3d} \gls*{nlte} \gls*{rt}
calculations remains a significant challenge. However, promising progress 
has been made in this regard \citep[e.g., ][]{Benedusi2023JCoPh}.

By extending the impact theory of pressure broadening 
\citep{Anderson1949PhRv,Baranger1958PhRv,Fiutak1962CaJPh}, 
the frequency redistribution function for a polarized atomic system was derived 
by \citet{Omont1972ApJ,Omont1973ApJ} and \citet{Domke1988ApJ}.
\citet{Bommier1997aA&A,Bommier1997bA&A} 
advanced the theoretical framework mentioned in \sref{sec3.3} 
by including higher-order terms in the perturbative expansion 
of the atom-radiation interaction, under the assumption of a two-level 
atom model with an unpolarized and infinitely sharp lower level. 
Later, quantum interference between $J$-states within the same atom term 
was incorporated into the frequency redistribution function for a 
two-term atom
\citep{Smitha2011ApJ,Smitha2013JQSRT,Bommier2017A&A}.\footnote{Note that
a multi-term atomic model accounts for quantum interference between
$J$-levels within the same term, whereas a multi-level atomic model does not.}
Independently, using a different theoretical approach, \citet{Casini2014ApJ}, 
\citet{Casini2016ApJ}, and \citet{Casini2017aApJ,Casini2017bApJ} derived 
the frequency redistribution function and corresponding expression for the emissivity
for $\Lambda$-type atomic systems. 
Applying the quantum theories proposed by 
\citet{Bommier1997aA&A,Bommier1997bA&A,Bommier2017A&A} 
and \citet{Casini2014ApJ}, \citet{AlsinaBallester2016ApJL,AlsinaBallester2022A&A} and 
\citet{delPinoAleman2016ApJL,delPinoAleman2020ApJ} independently
developed \gls*{1d} \gls*{rt} codes capable of
accounting for the Hanle and Zeeman effects, \gls*{prd} 
effects, as well as atomic polarization and $J$-state interference. Recently,
\cite{Riva2025A&A} demonstrated that both approaches yield 
identical polarization profiles for the Mg {\sc II} h and k lines when modeled with
a two-term atomic model.

To reduce computational cost, the angle-average approximation is commonly 
employed when solving the \gls*{rt} problem with \gls*{prd} effects. 
This approach decouples the angular and frequency dependencies of the 
redistribution functions by averaging the redistribution matrix over 
all directions \citep{Mihalas1978,Leenaarts2012A&A,Belluzzi2014A&A,AlsinaBallester2017ApJ}, 
thereby significantly accelerating the calculation compared with the 
general angle-dependent treatment. 
For the polarized case, the averaging of the redistribution matrix   
is performed excluding the scattering phase matrix \citep{Rees1982A&A}.
However, it is important to note that the this approximation can significantly 
impact the resulting polarization for some spectral lines, often around the line 
cores in the linear polarization profiles 
\citep{Nagendra2011A&A,Sampoorna2017ApJ,Janett2021A&A,Guerreiro2024A&A,Belluzzi2024A&A}. 

For specific spectral lines, the angle-average treatment of \gls*{prd} 
effects remains a valid approximation. 
For instance, \citet{Riva2024A&A} demonstrated that this
approximation provides reliable modeling of the Stokes profiles of the 
He {\sc II} Ly-$\rm \alpha$ at 30.4~nm. 
\citet{delPinoAleman2025ApJ} showed, using a \gls*{1d} semi-empirical 
atmospheric model with magnetic fields, that angle-dependent effects can 
lead to measurable differences in the linear polarization
near the core of the Mg {\sc II} k line.
Nonetheless, these authors pointed out that the impact of the angle-averaged 
approximation in the Mg {\sc II} k line is considerably reduced at the spectral 
resolution and polarimetric accuracy of the CLASP2 observations.

\section{Forward modeling of the Mg {\sc II} 
h and k line polarization}\label{sec4}

The broad polarization pattern in the far wings of the Mg {\sc II} h and k 
lines was first pointed out by \citet{Auer1980A&A}, who performed a simplified 
\gls*{rt} calculation by assuming fully coherent scattering in the observer's 
frame, which is a reasonable approximation for the far wings of spectral lines. 
A more rigorous radiative transfer investigation of the linear polarization across the entire 
Mg {\sc II} h and k linear polarization, 
in the absence of magnetic fields, was later carried out by 
\citet{Belluzzi2012ApJL}, using a two-term atomic model. 
Their results predicted an antisymmetric shape of the $Q/I$ profile around the 
center of the h line, and two negative troughs to the sides of a positive $Q/I$ peak
at the center of the k line.
These theoretical predictions were subsequently confirmed by the high-precision 
spectropolarimetric observation acquired by the CLASP2 sounding 
rocket experiment \citep[see][]{Rachmeler2022ApJ}.

\begin{figure*}[htp]
  \center
  \includegraphics[width=1.\textwidth]{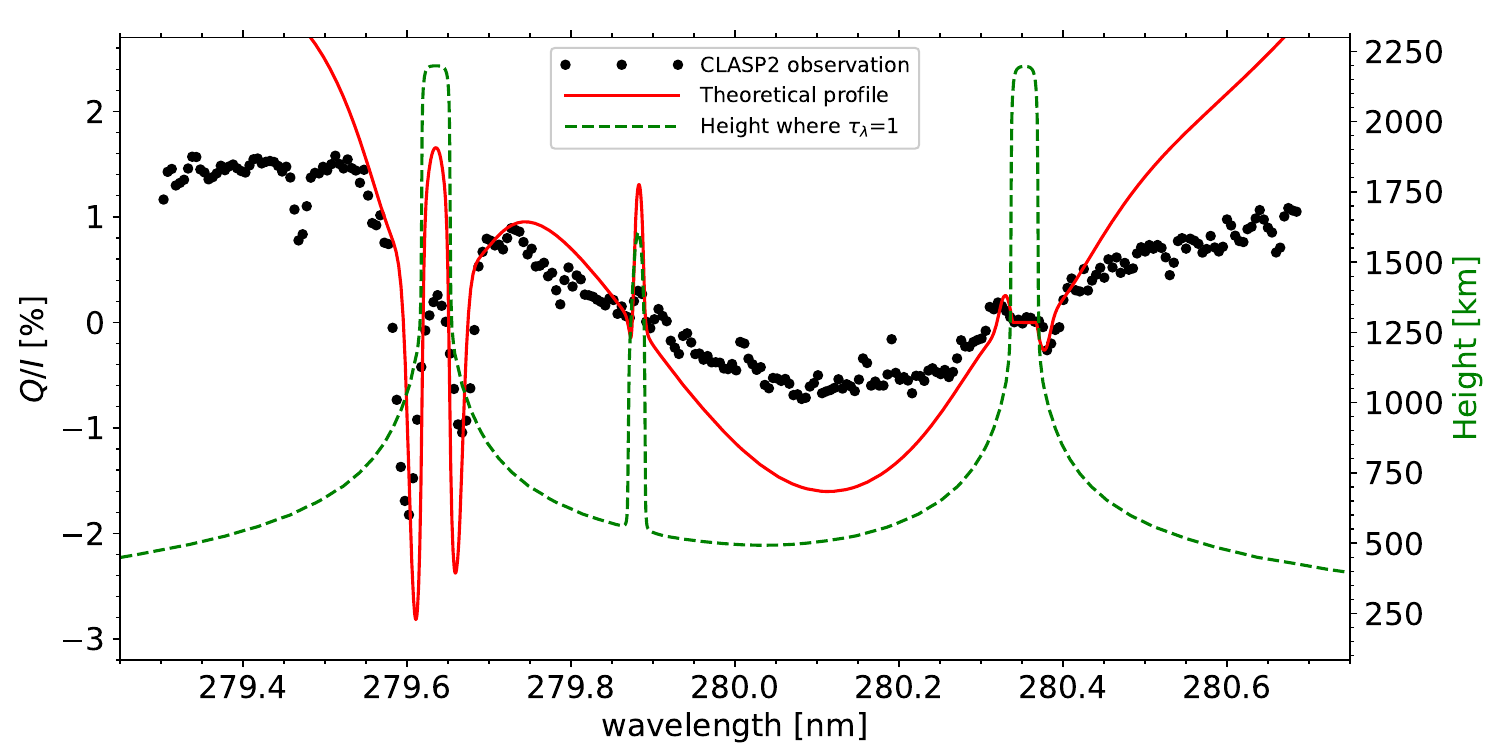}
  \caption{Temporally and spatially averaged $Q/I$ profile at $\mu=0.1$ observed by 
  CLASP2 (black dots), $Q/I$ profile resulting from the forward synthesis in the 
  FAL-C semi-empirical model with a three-term Mg {\sc II} atomic model in the absence 
  of a magnetic field (red curve), and height where the optical depth at each 
  wavelength is unity for $\mu=0.1$ (green dashed curve), corresponding to the scales 
  on the right axis.
  The reference direction for positive Stokes $Q$ is the parallel to the nearest 
  solar limb.
  The CLASP2 observational data are reproduced with permission from \citet{Rachmeler2022ApJ}.}
  \label{fig3}
\end{figure*}

\begin{figure*}[htp]
  \center
  \includegraphics[width=0.85\textwidth]{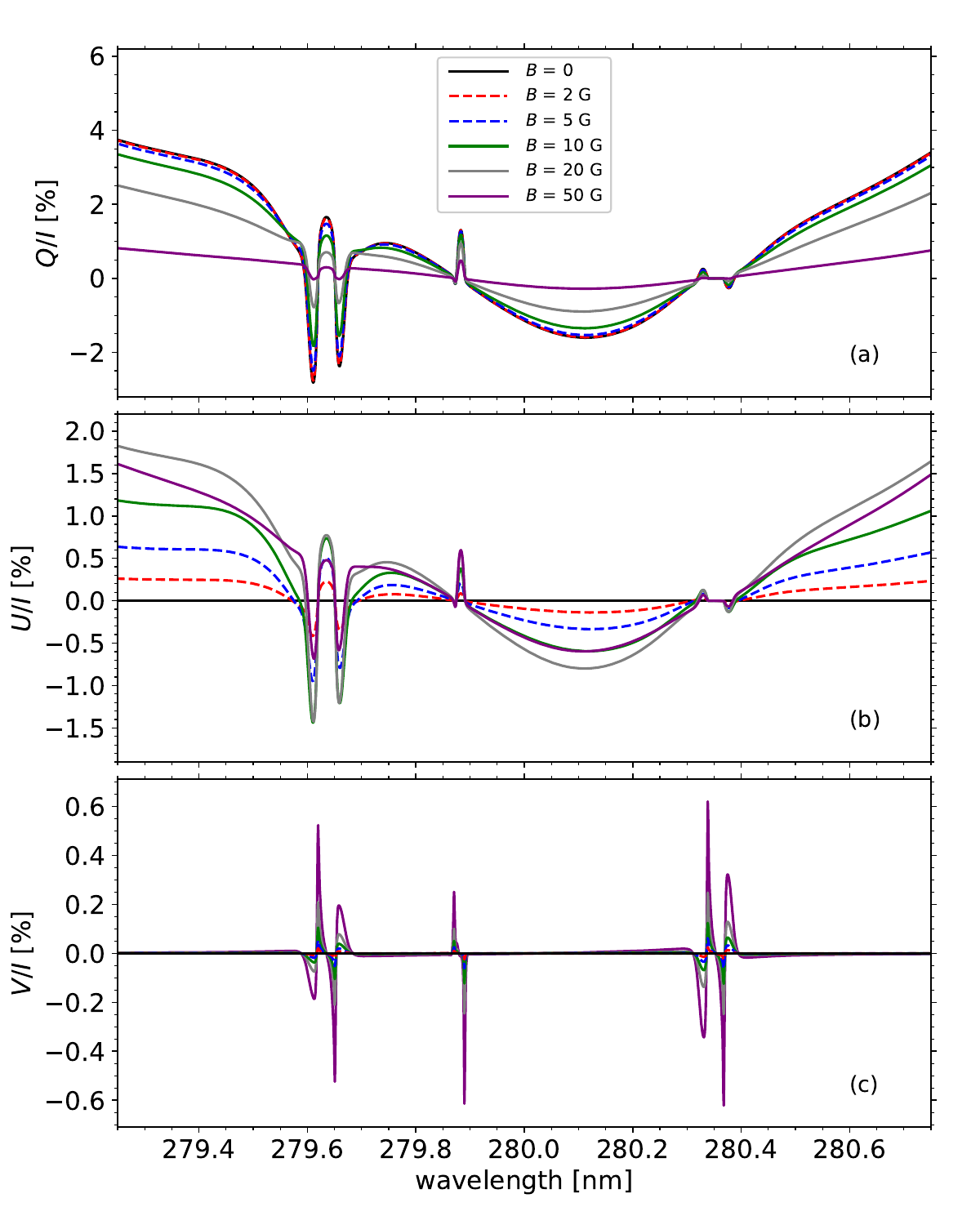}
  \caption{Synthetic $Q/I$ (panel a), $U/I$ (panel b), and
  $V/I$ (panel c) Mg {\sc II} h and k profiles in the FAL-C semi-empirical model 
  for a \gls*{los} with $\mu=0.1$.
  The different color curves correspond to calculations with magnetic
  fields inclined 45$^\circ$ with respect to the local vertical toward
  the observer, with different strengths indicated by the color (see legend in panel a). 
  The asymmetry in the $V/I$ profile at 279.88~nm 
  arises because it actually consists of two blended spectral components.
  This figure is similiar to that in \citet{delPinoAleman2020ApJ}, but 
  with a magnetic field with a different direction and strength.}
  \label{fig4}
\end{figure*}

The black dots in \fref{fig3} show the temporally and spatially averaged 
$Q/I$ profiles observed by CLASP2 near the solar limb.
Both the Mg {\sc II} h and k lines, as well as the subordinate lines 
at 279.88~nm were observed. 
The red curve shows the forward synthesis with HanleRT-TIC
in the C model 
of \citet[][here after FAL-C model]{Fontenla1993ApJ} using a three-term 
atomic model including five levels of Mg {\sc II} and the ground level 
of Mg {\sc III} (see also Figure 7 of \citealt{TrujilloBueno2022ARA&A}).
The green dashed curve indicates the height where the optical depth at each
wavelength equals unity, providing a rough estimate of the formation heights.

The observed $Q/I$ profile shows a positive signal at the center of 
the Mg {\sc II} k line, with negative troughs on either side.
As expected, the center of the Mg {\sc II} h line exhibits no $Q/I$ signal,  
because both the upper and lower levels have total angular momentum $J=1/2$ 
and thus cannot carry atomic alignment.
Consistent with theoretical predictions \citep[see][]{Belluzzi2012ApJL}, 
the antisymmetric $Q/I$ 
signal around the center of the h line, caused by \gls*{prd} effects
and $J$-state interference, was also detected by CLASP2. 
The polarization signals in the far wings, particularly the 
negative $Q/I$ signal in the region 
between the h and k lines, are clearly confirmed as well.

Although the synthesized profiles in \fref{fig3} reproduce the overall 
shape of the observed linear polarization, there are clear differences in both 
the width and amplitude of the profile. 
Part of these differences are due to the instrument's \gls*{psf}, 
which is not considered in the theoretical profiles, as well as to 
that the FAL-C model is an idealization of the real solar atmosphere. 
Nevertheless, the main reason for the difference in the amplitude of the $Q/I$ 
signals is that this calculation was performed without including magnetic fields. 
The presence of a magnetic field leads to depolarization of the $Q/I$ profiles, 
as illustrated in panel (a) of \fref{fig4}. 
The amplitude variation of the polarization demonstrates  
their valuable diagnostic potential for inferring the 
magnetic field properties of the chromospheric plasma.

The magnetic sensitivity of the Mg {\sc II} h and k lines has been studied
using a two-level atomic model focusing only on the k 
line \citep[][]{AlsinaBallester2016ApJL}, a two-term atomic 
model \citep{delPinoAleman2016ApJL,MansoSainz2019ASPC}, 
and a three-term atomic model including the subordinate 
lines \citep[][]{delPinoAleman2020ApJ}. 
Figure~\ref{fig4} illustrates the Stokes profiles of the Mg {\sc II} h and k lines, 
as well as the subordinate lines, synthesized in the FAL-C semi-empirical model 
for a \gls*{los} with $\mu=0.1$. 
The magnetic fields are inclined by 45$^\circ$ 
toward the observer with respect to the local vertical, with strengths 
ranging from 0 to 50~gauss.
The impact on the linear polarization at the center of the k line is dominated by 
the Hanle effect, which also operates on the subordinate lines.
In the wings, the magnetic sensitivity of the linear polarization
signals arises from \gls*{mo} effects.
The Stokes $V$ profiles of the h and k lines, primarily caused by the Zeeman 
effect, exhibit four lobes. The inner lobes originate in the upper chromosphere, 
while the outer lobes originate at relatively lower chromospheric heights.

\begin{figure*}[htp] 
  \center 
  \includegraphics[width=0.85\textwidth]{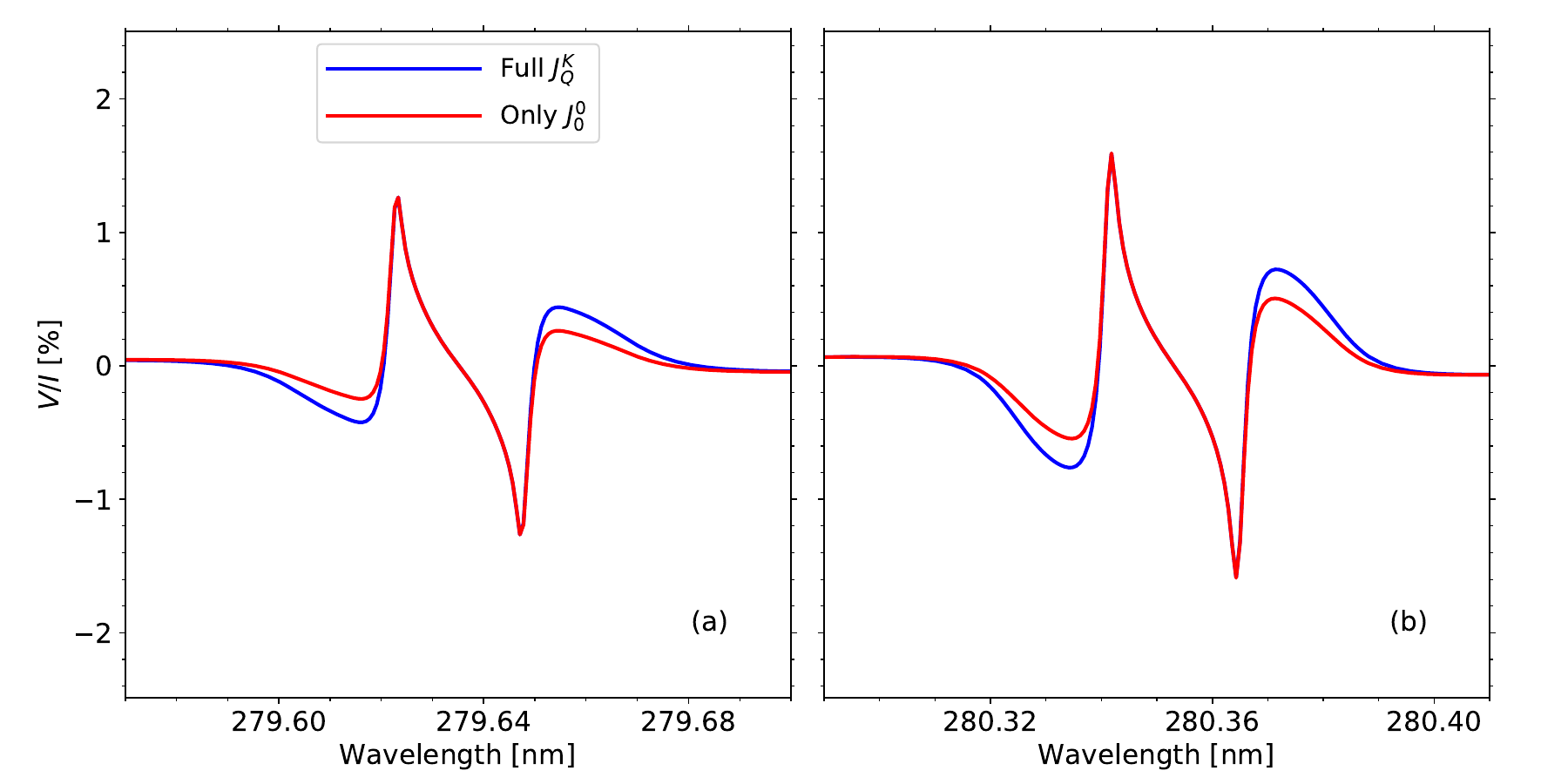} 
  \caption{$V/I$ profiles of the Mg {\sc II} k (panel a) and h (panel b) lines 
  in the FAL-C semi-empirical model for a \gls*{los} with $\mu=1.0$ and 
  for a magnetic field of 100~G parallel to the \gls*{los}. 
  The blue (red) curves correspond to the calculation including (neglecting) scattering 
  polarization. 
  This figure is similar to that in \citet{AlsinaBallester2016ApJL}.} 
  \label{fig5} 
\end{figure*} 

The outer lobes of the circular polarization are also impacted by the 
joint action of \gls*{prd} effects and the radiation field anisotropy
\citep{AlsinaBallester2016ApJL,delPinoAleman2016ApJL}. 
Figure~\ref{fig5} shows that accounting for the radiation field
anisotropy results in an enhancement of the signal of the outer 
lobes of the Stokes $V$ profiles. 
This enhancement is present even in the Mg {\sc II} h line, despite the 
fact that {its levels} cannot carry atomic alignment. 
If the radiation field anisotropy and the atomic polarization are neglected in an
inversion, these enhanced 
outer lobes will lead to an overestimation of the inferred longitudinal 
magnetic field \citep{Li2022ApJ}. 

\section{Non-LTE Stokes inversion}\label{sec5}

Inversion techniques are our most powerful tools to
extract the physical information encoded in spectral lines 
\citep{delToroIniesta2016LRSP,Reardon2023BAA}. 
The most advanced inversion codes can retrieve the stratification of 
magnetic field, temperature, electron density, gas pressure, and \gls*{los} 
and microturbulent velocities from spectropolarimetric observations, usually
under the assumption of hydrostatic equilibrium. 

The SIR code \citep{RuizCobo1992ApJ} was the first inversion tool 
capable of recovering a stratified model atmosphere from spectropolarimetric 
observations. 
Although SIR assumes \gls*{lte} and is thus restricted to photospheric 
lines, it made use of several important concepts, such as 
nodes, RFs, and the assumption of hydrostatic equilibrium to retrieve a
stratification of the gas pressure.  
These concepts have been adopted in subsequent depth-stratified inversion 
codes, such as SPINOR \citep{Frutiger2000A&A}, NICOLE 
\citep{Socas-Navarro2000ApJ,Socas-Navarro2015A&A}, 
SNAPI \citep{Milic2018A&A}, STiC \citep{delaCruz2016ApJL,delaCruz2019A&A}, 
DeSIRe \citep{RuizCobo2022A&A},
FIRTEZ \citep{PastorYabar2019A&A,Borrero2019A&A},
or HanleRT-TIC \citep{delPinoAleman2016ApJL,delPinoAleman2020ApJ,Li2022ApJ}. 

Inversion codes usually start with the synthesis of the line profiles 
from an initial guess of the atmospheric model. 
This model is parameterized using a prescribed number of nodes, from
which a fine stratification of the atmospheric parameters is interpolated. 
Iterative corrections to the model parameters at these nodes, based on RFs, are 
applied until the best fit to the observations is achieved. 
Typically, the Levenber-Marquardt algorithm \citep{NumericalRecipes} is 
employed to minimize a cost function.
Given that the inversion problem is inherently ill-posed, and there are often 
degeneracies between model parameters, regularization terms are usually included 
in the cost function. 
The regularization terms favor smooth stratifications of the model parameters 
over complex ones by employing penalties on these model parameters. 

The RFs describe the response (or change) of any given Stokes parameter 
of the emergent radiation at any given
wavelength to a perturbation of one of the model parameters at a specific location 
in the model atmosphere.
In some inversion codes, they can be computed analytically or semi-analytically,
as in SIR, SNAPI, FIRTEZ, and DeSIRe,
while in others they are calculated numerically, as in NICOLE, STiC, and HanleRT-TIC.
In the numerical approach, the RFs are obtained by perturbing the value at a single node 
and synthesizing the Stokes profiles in the resulting model atmosphere. 
This method is computationally expensive because it requires multiple 
spectral syntheses to obtain the RFs. 
Therefore, inversion codes following an analytical approach for the calculation of the RFs 
are usually faster than their numerical counterparts. 
However, their efficiency and accuracy have only been validated in the 
absence of atomic level polarization \citep{Milic2017A&A}. 
Moreover, deriving analytical RFs requires an explicit treatment 
of all interdependencies in the \gls*{se} equations, which becomes significantly 
challenging when including \gls*{prd} effects or atomic polarization. 

With the exception of HanleRT-TIC, all the above-mentioned inversion codes 
only take into account the 
Zeeman effect as polarization mechanism, and cannot handle most of the physical mechanisms
discussed in \sref{sec3}. 
A well-known exception not yet mentioned is the HAZEL code 
\citep{AsensioRamos2008ApJ}, which includes atomic level polarization as well as the 
Zeeman and Hanle effects, albeit under the assumption of an optically
thin and homogeneous slab model. 
HAZEL is commonly used for inverting spectropolarimetric observations 
of the He {\sc I} triplet at around 1083.0~nm
and the He {\sc I} D$_3$ lines \citep[e.g.][]{OrozcoSuarez2014A&A,MartinezGonzalez2015ApJ,EstebanPozuelo2025A&A}. 
Although HAZEL is mainly applied to prominence and filament observations, 
it has also been applied to other regions such as plages \citep{Anan2021ApJ} and spicules 
\citep{Centeno2010ApJ,MartinezGonzalez2012ApJ}. 
Nevertheless, the HAZEL code is not suitable for other chromospheric lines formed 
in optically thick plasmas, such as the Mg {\sc II} h and k lines. 
Moreover, it does not account for \gls*{prd} effects. 

To infer the magnetic field vector from spectropolarimetric observations 
of the Mg {\sc II} h and k lines, \citet{Li2022ApJ} developed the \gls*{tic}, 
which employs the HanleRT synthesis code 
\citep{delPinoAleman2016ApJL,delPinoAleman2020ApJ} as its forward modeling engine. 
HanleRT (the forward modeling module) and TIC (the Stokes inversion module) are  
now referred to as the HanleRT-TIC\footnote{HanleRT-TIC is 
publicly available at \url{https://gitlab.com/TdPA/hanlert-tic}.} radiative transfer code. 
HanleRT-TIC is a \gls*{1d} \gls*{nlte} spectral synthesis and inversion code capable of accounting 
for all the physical mechanisms introduced in \sref{sec3}, which are 
critical for modeling the polarization of the Mg {\sc II} h and k lines. 
To solve the system of linear equations for obtaining the corrections to the model
parameters from the numerically calculated Hessian matrix,
the code uses the modified \gls*{svd} method used by \cite{RuizCobo1992ApJ}. 
Moreover, electron and hydrogen number densities are computed under 
the assumption of \gls*{lte} by solving the equation of state using the 
method of \citet{Wittmann1974SoPh}, as implemented in the SIR code. 

The uncertainties of the model parameters at each node are estimated 
following the approach described in \citet{SanchezAlmeida1997ApJ} 
and \citet{delToroIniesta2003}, 
\begin{equation}\label{eq7} 
  \sigma_p^2 = \frac{\Delta\chi^2}{2N}H^{-1}, 
\end{equation}
where $N$ is the number of nodes, $\Delta\chi^2$ is the cost function 
excluding the regularization term, and $H^{-1}$ is the inverse 
of the Hessian matrix. 
In practice, $H^{-1}$ is often approximated by the inverse of the 
diagonal elements of the Hessian matrix. 
Since these diagonal elements are, in fact, the squares of the RFs, 
the uncertainties can be estimated directly from the RFs 
themselves. 
Although this method does not provide precise uncertainty estimates, it 
provides a relative indication of how well each model parameter is constrained 
at a given node. 

A more robust approach to estimate the uncertainties of the model parameters 
is the Monte Carlo approach
\citep{WestendorpPlaza2001ApJ,SainzDalda2023ApJ,Li2024aApJ}. 
In this method, a number of different Stokes profiles are generated by 
adding random noise consistent with the observational uncertainties 
to the measured profiles. 
The inversion of all these profiles produces a statistical distribution 
of inferred model parameters, which in turn provides a more accurate 
representation of the uncertainties. 

\section{Weak-field approximation}\label{sec6}

In addition to \gls*{nlte} inversions, the \gls*{wfa} is often employed for 
a rapid estimation of the magnetic field from the Stokes profiles. 
The \gls*{wfa} is applicable when the Zeeman splitting is much smaller than 
the Doppler width. 
The \gls*{wfa} expressions  were derived by \citet{LandiInnocenti1973SoPh} 
using perturbation theory, and by \citet{Jefferies1989ApJ} via Taylor 
series expansion. 
These equations are given by \citep[see][]{LL04}, 
\begin{subequations}\label{eq8}
\begin{align}
  V(\lambda) &= -\Delta\lambda_B\bar{g}\cos\theta
    \left(\frac{\partial I}{\partial\lambda}\right). \label{eq8a} \\
  \tilde{Q}(\lambda_0) &= -\frac{1}{4}\Delta\lambda^2_B
    \bar{G}\sin^2\theta\left(\frac{\partial^2I}{\partial\lambda^2}\right)_{\lambda_0}. \label{eq8b} \\
  \tilde{Q}(\lambda_{\rm w}) &= \frac{3}{4}\Delta\lambda^2_B\bar{G}\sin^2\theta
    \frac{1}{\lambda_{\rm w}-\lambda_0}
    \left(\frac{\partial I}{\partial \lambda}\right)_{\lambda_{\rm w}}. \label{eq8c} \\
  \frac{U(\lambda)}{Q(\lambda)} &= \tan 2\chi. \label{eq8d}
\end{align}
\end{subequations}
where $\tilde{Q}$ indicates the Stokes $Q$ parameter in the reference system where
Stokes $U$ is zero, and $\bar{G}$ is a constant that depends on the quantum
numbers of the levels involved in the transition. 
\eref{eq8a} can be used to estimate the longitudinal component of the magnetic field. 
Eqs.\,(\ref{eq8b}) and (\ref{eq8c}) can be used to estimate the transverse component, 
and they are valid in wavelengths close to
the line center ($\lambda_0$) and in the line wings ($\lambda_{\rm w}$), respectively. 
Finally, the azimuth of the magnetic field can be inferred from \eref{eq8d}. 

Although the \gls*{wfa} assumes height-independent physical parameters 
in the formation regions of the spectral lines 
and neglects atomic polarization and \gls*{prd} effects, it enables 
a very fast estimation of the magnetic field. 
Note that \eref{eq8a}, the \gls*{wfa} for Stokes $V$, only requires 
that the longitudinal component of the magnetic field is constant in the 
formation region. 
The \gls*{wfa} has been successfully applied in numerous studies to 
extract the magnetic field vector from spectropolarimetric 
observations, \citep[e.g.,][]{MartinezGonzalez2009ApJ, 
MartinezGonzalez2012MNRAS,Morosin2020A&A,EstebanPozuelo2023A&A,
Schad2024SciA,QuinteroNoda2025A&A}. 

The reliability of the \gls*{wfa} for estimating the longitudinal 
magnetic field from the Mg {\sc II} h and k lines has been studied by 
\citet{delPinoAleman2016ApJL}, \citet{Centeno2022bApJ}, and 
\citet{AfonsoDelgado2023aApJ}. 
From spectropolarimetric syntheses performed
in the FAL-C atmospheric model, \citet{Centeno2022bApJ} found 
that applying the \gls*{wfa} to all four lobes of the Stokes $V$ 
profile results in an underestimation of the longitudinal magnetic 
field by approximately 13\%, whereas the values derived from only 
the inner lobes closely match the input magnetic field strength. 
Furthermore, spectral degradation to a resolution of 30~000, similar 
to that of CLASP2, increases the error to approximately 18\% and 10\% 
when derived from the four lobes and inner lobes, respectively.

\section{Interpretation of the Observations}\label{sec7} 

The intensity profiles of the Mg {\sc II} h and k lines have been routinely 
observed by the IRIS satellite since 2014, with a spectral sampling of 
25.4~m{\AA}\,pixel$^{-1}$. 
The \gls*{nlte} inversion of the intensity profiles of the Mg {\sc II} h 
and k lines can be performed using, for example, the STiC code, which is built upon the 
RH code \citep{Uitenbroek2001ApJ}, and is thus 
capable of accounting for \gls*{prd} effects and the Zeeman effect. 
Lately, \citet{SainzDalda2019ApJL,SainzDalda2024ApJS} 
built a database, dubbed IRIS$^2$, which contains representative intensity 
profiles of the Mg {\sc II} h and k lines observed by IRIS. 
The corresponding representative atmospheric models were obtained 
using the STiC inversion code. 
By matching observed profiles to this precomputed database, IRIS$^2$ 
enables a very quick estimation of the inferred model atmosphere,
with a reduction of computational cost of about
10$^5$ - 10$^6$, while maintaining an 
accuracy comparable to that of the full inversion. 

The Mg {\sc II} h and k lines provide constraints on the chromospheric 
temperature and turbulent velocity \citep{Vissers2019A&A,Bryans2020ApJ}. 
\citet{daSilvaSantos2020A&A} performed an inversion of IRIS observations of 
the Mg {\sc II} h and k lines together with radio observations 
at 1.25~mm from the Atacama Large Millimeter/submillimeter 
Array \citep[ALMA,][]{Wootten2009IEEEP} using the STiC code. 
Their results placed robust constraints on the temperature 
and turbulent velocity over a wide range of heights. 
Additionally, an inversion based on the IRIS$^2$ database, carried out 
by \citet{Bose2024NatAs}, revealed a strong correlation between the heating 
in plage and moss regions. 
These studies demonstrate the diagnostic potential of the 
Mg {\sc II} h and k lines for probing the thermodynamic properties 
in the chromosphere. 
For a comprehensive overview of the results achieved by IRIS, 
see \citet{DePontieu2021SoPh}.

\subsection{CLASP2 and CLASP2.1 observations}\label{sec7.1} 

In order to demonstrate the theoretically predicted diagnostic potential of 
the Mg {\sc II} h and k lines for studying chromospheric magnetic fields, 
the CLASP2 sounding rocket was launched on April 11, 2019, measuring the Stokes 
profiles of these lines. 
The observed spectral range spanned from 279.30 to 280.68~nm, covering the 
Mg {\sc II} h and k lines, the subordinate lines at 279.88~nm, and 
the Mn {\sc I} lines at 279.91 and 280.19~nm with a spectral sampling of 
49.9~m{\AA}\,pixel$^{-1}$. 
The \gls*{fwhm} of the instrument's profile, resulting from the convolution 
of the slit width with the spectral \gls*{psf}, is
110~m{\AA} \citep[][]{Song2018SPIE,Tsuzuki2020SPIE}. 

During the $\sim$5~min observation window, CLASP2 performed sit-and-stare 
observations using a 196'' long spectrograph slit positioned at an 
active-region plage and a quiet Sun region close to the limb. 
The spatial resolution along the slit direction was 0.53'' per pixel, 
and the polarization accuracy was better than 0.1\%. 

Motivated by the success of this suborbital mission, a reflight of 
CLASP2, dubbed CLASP2.1, was carried out on October 8, 2021. 
Instead of the sit-and-stare observation, CLASP2.1 scanned a 
two-dimensional field of view over an active region plage located 
near a sunspot with a raster step size of approximately 1.8''. 
CLASP2.1 measured the full Stokes parameters in the same spectral 
range as CLASP2. 
However, the polarimetric accuracy was slightly worse (around 
$10^{-3}$ at the intensity peaks of the Mg II h and k lines), 
due to the reduced integration time. 
For further details on the CLASP2 and CLASP2.1 missions and their observations, 
see \citet{Ishikawa2021SciA,Ishikawa2023ApJ}, 
\citet{Rachmeler2022ApJ}, 
\citet{TrujilloBueno2022ARA&A}, 
\citet{Li2024aApJ}, \citet{Song2025ApJ}, \citet{Ishikawa2025ApJ}
and \citet{AfonsoDelgado2025ApJ}.

\subsection{Inversion of the Stokes $I$ and $V$ profiles}\label{sec7.2} 

\begin{figure*}[htp]
  \center
  \includegraphics[width=0.7\textwidth]{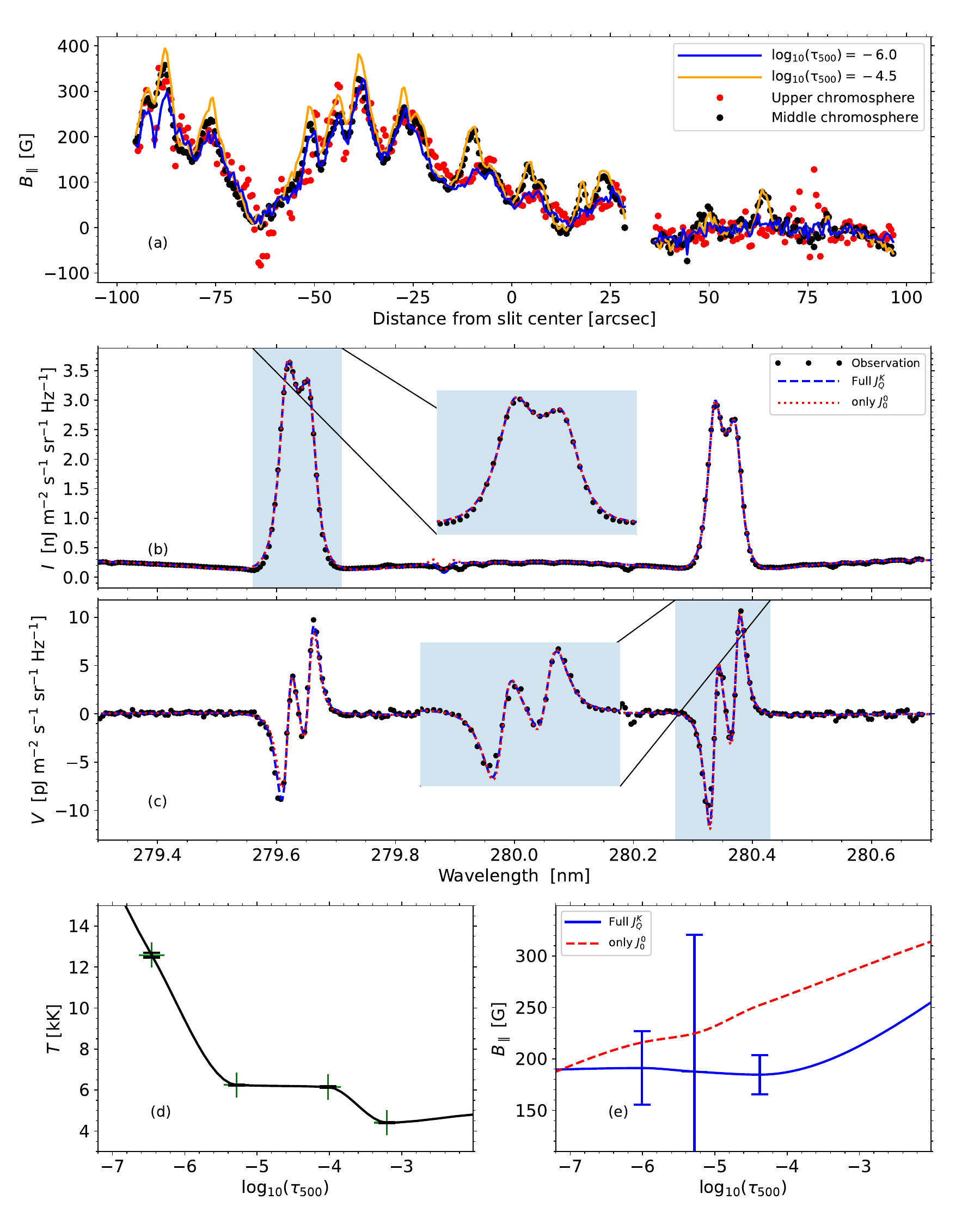}
  \caption{
  Longitudinal magnetic field inferred from the application of the \gls*{wfa} 
  to the inner (red dots) and outer (black dots) lobes of the $V/I$ profiles of the 
  Mg {\sc II} h and k lines obtained by CLASP2 \citep{Ishikawa2021SciA}.
  Longitudinal magnetic field inferred from the inversion of the
  intensity and circular polarization profiles of the Mg {\sc II} h
  and k lines at
  log$_{10}(\tau_{500})=-6.0$ (blue curve in panel a) and $-4.5$ (orange curve in panel a)
  from \citet{Li2023ApJ}.
  Panels (b) and (c) show the observed (black dots) and fit (dashed blue and
  dotted red curves) Stokes $I$ and $V$ 
  profiles for a pixel in the plage region.
  The dashed blue (dotted red) curves in panels (b) and (c)
  correspond to the inversion including (neglecting) the anisotropy terms.
  Panels (d) and (e) show the inferred temperature (black curve) and longitudinal component of the
  magnetic field from the inversion including (solid blue curve) and
  neglecting (dashed red curve) the anisotropy terms,
  for the same pixel, with error bars for the temperature and the
  magnetic field calculated including the anisotropy terms. The green ``+'' in
  panel (d) signal the position of the temperature nodes.
  Image reproduced with permission from \citet{Li2023ApJ}.} 
  \label{fig6}
\end{figure*}

Given the complexity associated with interpreting the linear polarization of the 
Mg {\sc II} h and k lines (see \sref{sec3}), \citet{Ishikawa2021SciA} first analyzed 
the Stokes $I$ and $V$ profiles using the \gls*{wfa}. 
By applying the \gls*{wfa} independently to the inner lobes of the Mg {\sc II} h and k 
lines, and to the outer lobes of the h line, they derived 
the longitudinal components of the magnetic fields in the upper 
(red circles in panel (a) of \fref{fig6}) and middle (black circles) 
chromosphere, respectively. 
Inside the magnetic flux concentrations, the inferred field strength 
reached up to 300~G in the middle and upper chromosphere. 
This value is consistent with the magnetic field strength reported by 
\citet{daSilvaSantos2023ApJL}, and is slightly lower than
those in \citet{Morosin2020A&A}, which are inferred from the 
Ca {\sc II} 854.2~nm by using the \gls*{wfa}. 
This discrepancy can be explained by that the observations were obtained 
from different target regions with different \gls*{los} directions, 
and that the Ca {\sc II} 854.2 nm line is sensitive to relatively lower atmospheric layers 
than the Mg {\sc II} h and k lines \citep{daSilvaSantos2018A&A}.

The same dataset was later analyzed by \citet{Li2023ApJ} using 
HanleRT-TIC. 
Since the circular polarization is not impacted by $J$-state 
interference, it was neglected in the inversion to reduce the computational cost. 
The resulting longitudinal magnetic field at log$_{10}(\tau_{500})=-6.0$ 
and $-4.5$, shown as the blue and orange curves in panel (a) 
of \fref{fig6}, respectively, closely match the results obtained 
using the \gls*{wfa}. 
An example of the inversion of the Stoke $I$ and $V$ profiles in the 
plage region is shown in panel (b) and (c), where the blue and red curves 
correspond to the fits from the inversions with and without the anisotropy 
terms, respectively. 
The temperature and the longitudinal magnetic field that resulted from the inversion 
are shown in panel (d) and (e), respectively. 
As seen in the figure, both inversion setups reproduce the observations well. 
As expected, the inversion without anisotropy terms required 
a stronger magnetic field in order to fit the circular polarization in the line 
wings, as discussed in \sref{sec4}. 
Although the impact of the anisotropy is expected to appear only 
in the outer lobes, primarily affecting the inference of the longitudinal 
magnetic field in the middle chromosphere, the spectral \gls*{psf} couples 
different wavelengths, resulting in an impact also on
the inferred magnetic field in the upper chromosphere. 

The longitudinal magnetic field inferred from the \gls*{wfa} exhibits a 
strong correlation with the electron pressure (the product of temperature 
and electron density) derived from the IRIS$^2$ 
database \citep{SainzDalda2019ApJL}, suggesting a magnetic origin of the 
chromospheric heating in active region plages \citep{Ishikawa2021SciA}. 
This correlation was further confirmed by the \gls*{nlte} inversion 
results \citep{Li2023ApJ}. 
However, a similarly significant correlation is not observed in the 
results obtained from the inversion of the He {\sc I} triplet at
1083.0~nm lines \citep{Anan2021ApJ}. 

\begin{figure*}[htp]
  \center
  \includegraphics[width=1.0\textwidth]{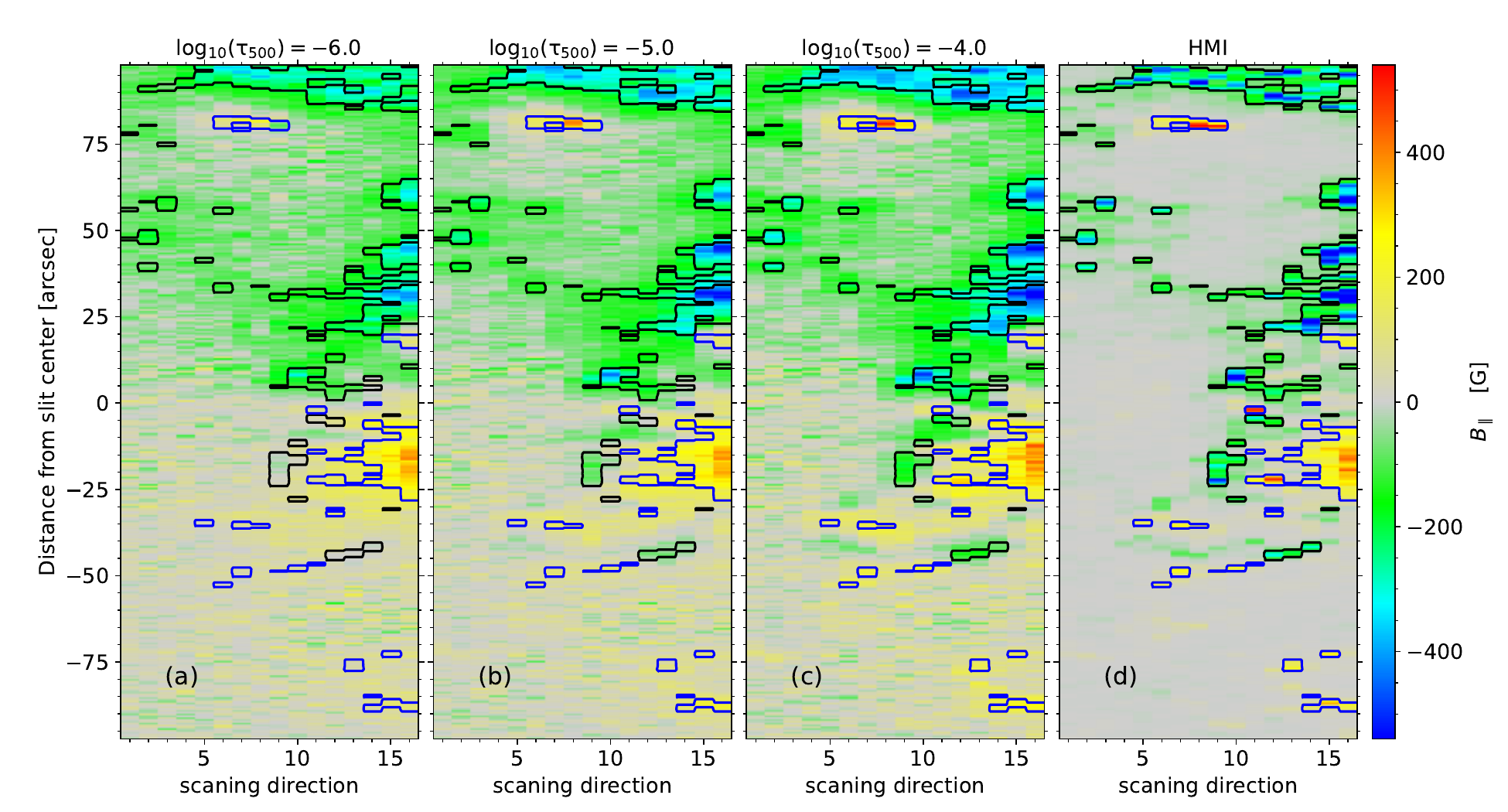} 
  \caption{Panels (a), (b), and (c) display the longitudinal magnetic 
  field at log$_{10}(\tau_{500})=-6.0$, $-5.0$, 
  and $-4.0$, respectively, resulting form the inversion of the CLASP2.1 observations. 
  Panel (d) shows the corresponding HMI magnetogram. 
  Blue and black contours indicate the HMI magnetic field strengths of 
  70~G and -70~G, respectively. 
  Image reproduced with permission from \citet{Li2024aApJ}.} 
  \label{fig7}
\end{figure*}

The CLASP2.1 mission enabled spatial scanning to capture a two-dimensional 
field of view of an active region plage located near a sunspot. 
Panels (a), (b), and (c) of \fref{fig7} show the longitudinal magnetic 
fields at log$_{10}(\tau_{500})=-6.0$, $-5.0$, and $-4.0$, respectively, 
obtained through pixel-by-pixel inversion of the Stokes $I$ and $V$ 
profiles \citep{Li2024aApJ}. 
The polarity of the magnetic field flux concentrations in the chromosphere 
is generally consistent with those found in the photosphere, as seen in
the HMI magnetogram shown in panel (d). 
These magnetic field flux concentrations expand with height due to the 
lower gas pressure in the chromosphere compared to the photosphere, and thus 
occupy larger areas in the chromosphere. 

The magnetic field obtained from the Stokes inversion shows strong 
correlation with the temperature 
and electron density in the chromospheric plage, as well as with the intensity of 
the AIA 171~{\AA} band in the overlying moss region \citep{Li2024aApJ}. 
The correlation between the intensities of spectral lines formed in the moss 
region and in the underlying chromospheric plage have been 
reported by \citet{Vourlidas2001ApJ}, \citet{DePontieu2003ApJ}, 
and \citet{Bose2024NatAs}, suggesting a possiblly common heating 
mechanism operating in both the chromospheric plage and the transition 
region moss. 
However, \citet{Judge2024ApJ} reported an absence of such a correlation between 
the AIA 171~{\AA} intensity and the chromospheric magnetic field inferred 
from the Ca {\sc II} 854.2~nm line using \gls*{wfa} 
in a plage region that includes the footpoints of coronal loops, 
even though the heating appears to be concentrated around unipolar 
chromospheric magnetic field regions. 

Overall, the polarities of the longitudinal magnetic fields in
the upper and middle chromosphere obtained by CLASP2.1 
are consistent. 
However, in certain regions, polarity changes with height have been reported 
by \citet{Li2024aApJ} and \citet{Ishikawa2025ApJ}. 
To verify the reliability of the polarity change, a Monte Carlo simulation was 
employed by \citet{Li2024aApJ}, confirming that the inferred field strengths 
exceed the uncertainties caused by the noise. 
\citet{Song2025ApJ} reported a coronal loop brightening in the vicinity 
of the polarity change region, suggesting a possible connection between 
magnetic field geometry and coronal activity. 
Similar polarity reversals between the photosphere and chromosphere have also 
been reported by \citet{Mathur2023ApJ} based on spectropolarimetric 
observations of photospheric and chromospheric lines. 
Even in the photosphere, polarity reversals can also been detected with 
the Fe {\sc I} 630.1~and 630.2~nm lines \citep{Liu2025A&A}.

\subsection{Full Stokes inversion}\label{sec7.3}

\begin{figure*}[htp]
  \center
  \includegraphics[width=0.8\textwidth]{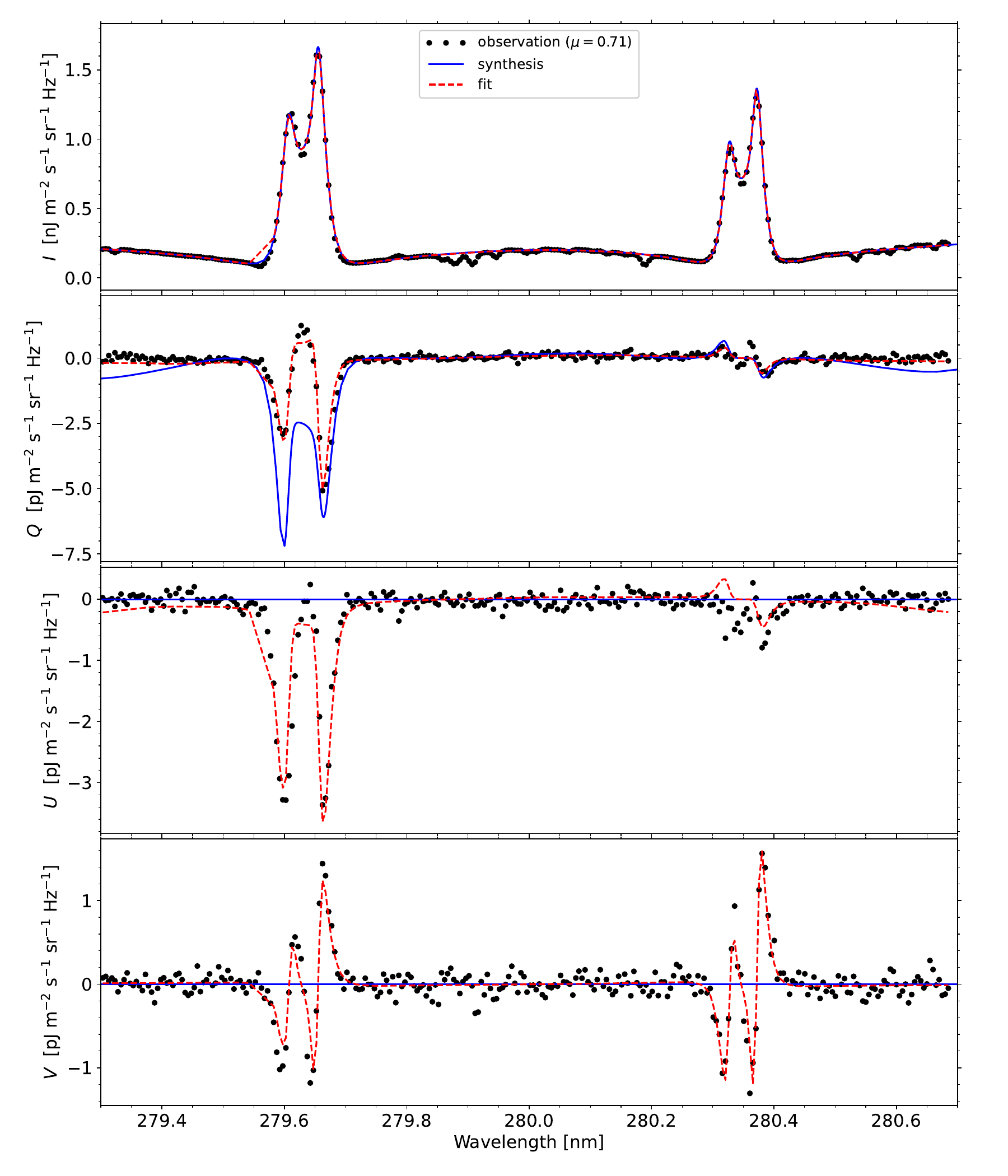}
  \caption{From top to bottom, Stokes $I$, $Q$, $U$, and $V$, respectively, of the Mg {\sc II} k 
  and h lines in a pixel of the plage region with $\mu=0.71$ acquired by CLASP2 (black dots). 
  The blue curves correspond to synthetic Stokes profiles calculated in the 
  non-magnetized model atmosphere resulting from the inversion of the intensity profile. 
  The red curves show the full Stokes inversion results, including the parameterization 
  of the radiation field tensor components ${J^\dagger}^K_Q$ and the magnetic field. 
  Note that the $k_{1v}$ minimum is blended with a Mn {\sc I} line at 
  around 279.57~nm; thus the affected wavelength points were excluded from the inversion
  and from the plot, the reason why there is a straight line in that spectral region for the
  intensity. 
  Image reproduced with permission from \citet{Li2024bApJ}.} 
  \label{fig8}
\end{figure*}

In \gls*{1d} atmospheric models without horizontal macroscopic velocities, 
the linear polarization can only be either parallel or perpendicular to the nearest 
solar limb, unless there is an inclined 
magnetic field. Under such circumstances, when either of these directions 
is chosen as the reference for linear polarization, 
Stokes $U$ can only arise in the presence of a magnetic field.
However, in the real solar atmosphere, horizontal inhomogeneities in the 
plasma temperature and density, as well as the gradients of the horizontal 
components of the macroscopic velocity, are able to break the axial symmetry. 
As a result, Stokes $U$ can appear even in the absence of a magnetic 
field \citep{MansoSainz2011ApJ,Stepan2016ApJL,JaumeBestard2021ApJ}. 

The black dots in \fref{fig8} display the Stokes profiles observed by CLASP2 
in a pixel within the plage region. 
The blue curves represent the Stokes profiles synthesized in the \gls*{1d} 
non-magnetized model atmosphere obtained by inverting the intensity profile. 
As expected, the blue curves show zero signals in Stokes $U$ and $V$, 
while there is a clear Stokes $Q$ signal due to scattering polarization.
Notably, the amplitude of the right trough of the k line in this synthetic 
Stokes $Q$ profile closely matches that of the observation. 
Given that the Hanle and \gls*{mo} effects typically 
depolarize and rotate the linear polarization (transforming Stokes $Q$ into $U$ 
in this case), it is thus not possible to find a magnetic field vector 
that simultaneously reproduces all the observed Stokes $Q$, $U$, and $V$ profiles. 
This highlights the limitations of a \gls*{1d} plane-parallel modeling 
and suggests that horizontal \gls*{rt} plays a significant role.

\subsubsection{Parameterization of the lack of axial symmetry}\label{sec7.3.1}


In addition to the physical mechanisms mentioned in \sref{sec3}, 
axial-symmetry breaking caused by horizontal inhomogeneities and \gls*{rt}, i.e. 
\gls*{3d} effects, must be taken into account when inverting the 
full Stokes profiles of the Mg {\sc II} h and k lines. 
However, the development of such an inversion code that simultaneously accounts 
for \gls*{prd} effects and $J$-state interference in full
\gls*{3d} remains a significant challenge. 

In a \gls*{1d} \gls*{rt} calculation, the model atmosphere is assumed to be plane-parallel,
with axially symmetric plasma thermodynamic properties.
As a result, in a non-magnetic and static \gls*{1d} model the radiation field
tensor components $J^2_Q$ with $Q\ne 0$ vanish \citep[for a detailed description of 
the $J^K_Q$ tensor, refer to][]{LL04}. 
However, in a \gls*{3d} atmosphere, where horizontal \gls*{rt} is taken 
into account, these components may be non-zero even in the non-magnetic and static case. 
There is thus a missing contribution to these tensor components
in a \gls*{1d} model atmosphere, which can lead to significant 
inaccuracies in the inversion of the linear polarization of the Mg {\sc II} 
h and k lines. 

\citet{Li2024bApJ} proposed a method to parameterize this missing
contribution and implemented it in the HanleRT-TIC code by introducing ad-hoc radiation 
field tensor components. 
Specifically, they define the radiation field tensor components 
as, 
\begin{subequations}\label{eq9}
  \begin{align}
  J'^2_1 & = J^2_1 + {J^\dagger}^2_1, \label{eq9a} \\ 
  J'^2_2 & = J^2_2 + {J^\dagger}^2_2, \label{eq9b} 
  \end{align}
\end{subequations}
where $J^2_1$ and $J^2_2$ are the tensor components obtained from 
standard \gls*{1d} \gls*{rt} calculations. 
${J^\dagger}^2_1$ and ${J^\dagger}^2_2$ are the ad-hoc parameters 
that mimic the missing contributions due to \gls*{3d} effects. 
They are described by four free parameters in the inversion, since the 
tensor components are complex numbers. 
The resulting $J'^2_1$ and $J'^2_2$ are then used in the \gls*{se} 
equations and in the computation of the emissivity for the \gls*{rt} equations. 

The red curves in \fref{fig8} show the results obtained by including these ad-hoc radiation 
field tensor components as free parameters. 
This approach achieves an excellent fit to the observed Stokes profiles. 
A magnetic field strength on the order of tens of gauss, decreasing 
with height, is inferred from the selected plage region pixel \citep{Li2024bApJ}. 
However, due to the high computational cost of the inversion, potentially
requiring hundreds of CPU-hours just for a single pixel, 
this method {would have to be improved} for application to large datasets.
 
\subsubsection{Vertical gradients in the horizontal velocity}\label{sec7.3.2}

\begin{figure*}[htp]
  \center
  \includegraphics[width=1.0\textwidth]{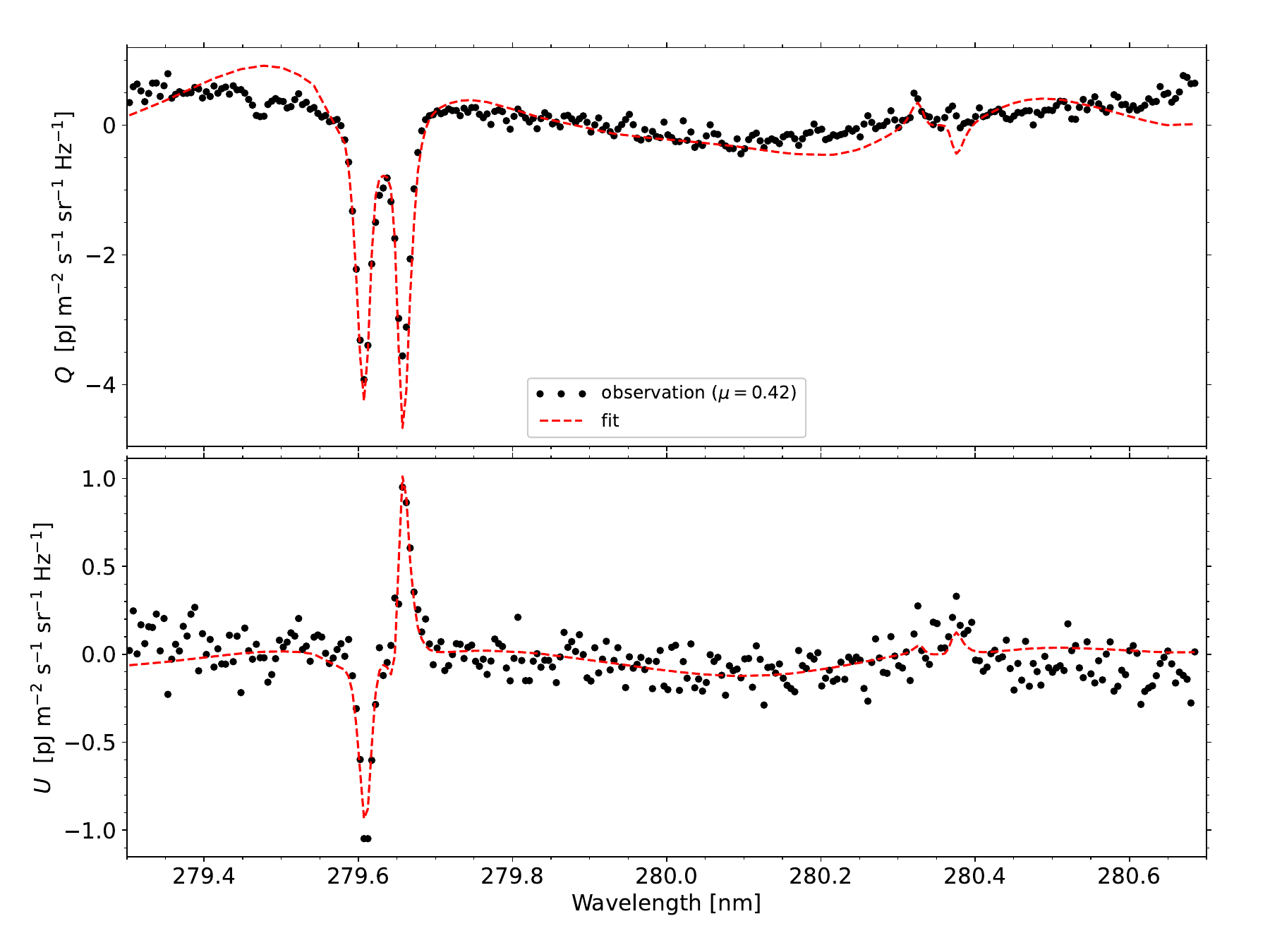} 
  \caption{Stokes $Q$ (top panel) and $U$ (bottom panel) profiles for
  the observation (black dots) and the best inversion fit (red curve)
  for one pixel in CLASP2 showing an antisymmetric $U$ signal.
  The inversion needs to take into account the 
  gradients of the horizontal components of the macroscopic velocity. 
  Image reproduced with permission from \citet{Li2024bApJ}.} 
  \label{fig9}
\end{figure*}

Several pixels in quiet regions observed by CLASP2 showed antisymmetric 
Stokes $U$ signals around the center of the k line. 
In a \gls*{1d} model atmosphere, such profiles can only be synthesized by 
introducing a vertical gradient in the horizontal component of the 
macroscopic velocity \citep{Li2024bApJ}. 
\fref{fig9} shows the observed linear polarization profiles in one such 
pixel from the quiet Sun observation for a \gls*{los} with $\mu=0.42$, 
as well as the corresponding fit from the inversion.
A horizontal velocity difference of about 5~$\rm km\,s^{-1}$ between the 
upper and lower chromosphere was inferred from this inversion. 
Antisymmetric Stokes $U$ profiles were detected in only a few pixels, 
suggesting that such horizontal velocity gradients may not be significant
in most regions of the quiet solar chromosphere. 

\subsubsection{Degeneracies and ambiguities}\label{sec7.3.3}

Degeneracies among the ad-hoc radiation tensor components
${J^\dagger}^K_Q$ and the magnetic field vector has been reported 
by \citet{Li2024bApJ}. 
As it happens with other degeneracies, changes in some of the parameters
can be partially compensated with changes in others. However, it is
important to emphasize that there is some degree of degeneracy, but
not complete degeneracy. For instance, the longitudinal component of the
magnetic field modulates both Stokes $V$, the coupling between Stokes $Q$ and
$U$ in the \gls*{rt} equations, and the depolarization of the linear
polarization of the line wings. Regarding the ad-hoc radiation tensor
components, they are added to the proper tensor components resulting from
the \gls*{1d} calculations in each iteration of the forward modeling. 
Although they get ``mixed'' with the anisotropy calculated in the vertical
reference frame when there is a non-vertical magnetic field, the
solution of the \gls*{1d} \gls*{rt} problem must be self-consistent 
and physical, and produce emergent profiles that fit the observation.
In summary, the physics of the problem imposes some constraints on the
mentioned degeneracy. 
Thus, not all combinations of these parameters are possible, although 
it is very difficult to demonstrate it formally
when \gls*{rt}, scattering polarization, \gls*{prd} effects, and the
Hanle effect are accounted for.

An alternative method to constrain the degeneracy is to estimate 
the ad-hoc radiation tensor components from the intensity map, as 
demonstrated by \citet{Zeuner2020ApJL,Zeuner2024ApJ} in their 
investigations on the scattering polarization of the Sr {\sc I} 
4607~{\AA} line. However, while this approach may be effective for 
photospheric lines, it does not yield good results for chromospheric lines.

It is important to mention that the ad-hoc radiation tensor components
are not intrinsic parameters in the \gls*{rt} calculations. 
They are exclusively introduced to mimic the effect of horizontal \gls*{rt},
which cannot be taken into account in \gls*{1d} inversion codes. 
These contributions can, in principle, be fully accounted for through 
full 3D RT calculations  \citep{Stepan2013A&A}. 
A 3D inversion framework has recently been developed by \citet{Stepan2022A&A} 
and applied to the synthesized Stokes profiles of the Mg {\sc II} k 
line in a prominence model \citep{Stepan2024A&A}. 
Currently, this approach is limited to \gls*{crd}, but there is an
ongoing effort to extend it to include coherent scattering in the
line wings and to apply it to the CLASP2.1 observations.

Regarding the ambiguities in the direction of the magnetic field vector,
in contrast to the well known 180$^\circ$ azimuthal ambiguity of the Zeeman 
effect \citep{Metcalf2006SoPh}, the ambiguity suffered by the Hanle 
effect is associated with the magnetic field vector in the local vertical 
reference frame \citep[see the Hanle diagrams in][]{LL04}. 
In the saturated regime of the Hanle effect, it is known as the Van Vleck 
ambiguity, which leads to multiple possible solutions for the magnetic field 
vector (see \citealt{AsensioRamos2008ApJ} for the He {\sc I} triplet 
at around 1083.0~nm, and \citealt{Casini2017SSRv} for the forbidden 
coronal lines). 
However, in the case of the Mg {\sc II} k line, the ambiguity is more complex 
due to the significant influence of the \gls*{mo} and \gls*{prd} effects on 
the linear polarization, particularly in the wings. 
This complexity has been demonstrated through the inversions 
by \citet{Li2024bApJ}, whose results show that different 
combinations of magnetic field inclination and azimuth can produce 
very similar Stokes $Q$ and $U$ profiles, successfully fitting the CLASP2 
observations.

\section{Summary and future perspectives}\label{sec8}

The polarization signals of the Mg {\sc II} h and k lines have demonstrated 
significant potential for diagnosing the magnetic field in the solar chromosphere, 
which is key to address some of the still open questions in solar physics. 
The CLASP2 and CLASP2.1 sounding rocket experiments successfully acquired  
spectropolarimetric observations of these lines. 
Analyses built on these unprecedented data have demonstrated the 
capability to infer magnetic fields from such observations 
\citep{Ishikawa2021SciA,Li2023ApJ,Li2024bApJ,Li2024aApJ,
Song2025ApJ,Ishikawa2025ApJ,AfonsoDelgado2025ApJ}. 

The \gls*{wfa} provides a fast estimation of the longitudinal component of the 
magnetic field, while \gls*{nlte} inversion techniques can retrieve the magnetic 
field vector from the full Stokes profiles. 
However, the \gls*{nlte} inversions require accounting for a plethora of physical 
ingredients which make them computationally heavy.

To accelerate the inversion process, several approaches have been proposed. 
These include database-based inversions \citep{SainzDalda2019ApJL}, 
convolutional neural networks for rapid computation of RFs 
\citep{Centeno2022aApJ}, or graph networks for the prediction of
departure coefficients \citep{VicenteArevalo2022ApJ}. 
For a comprehensive review on machine learning applications in Stokes 
inversions, see \citet{AsensioRamos2023LRSP}.  

Currently, \gls*{nlte} inversions of the Mg {\sc II} h and k lines 
are limited to \gls*{1d} atmospheric models. 
To reproduce the observed Stokes profiles, ad-hoc radiation tensor components need to 
be introduced in order to mimic the contribution from the \gls*{3d} effects. 
These ad-hoc tensor components introduce degeneracies among model parameters. 
Overcoming this limitation requires a fully \gls*{3d} inversion framework. 
At present, such a framework has been developed only under the assumption of \gls*{crd},  
though efforts are underway to partially extend this approach. 
Moreover, neural fields offer a promising way to represent model parameters in 
the \gls*{3d} inversion \citep{AsensioRamos2023SoPh,DiazBaso2025A&A}.

Finally, we emphasize the potential of future space missions with CLASP2-like 
capabilities to advance our understanding of magnetic fields in the solar chromosphere. 
Moreover, in addition to the lines present in the CALSP2/2.1 spectral 
range \citep{AfonsoDelgado2025ApJ}, several Fe {\sc II} lines are located in the 
wavelength range between 250 and 278 nm \citep{Judge2021ApJ}. 
Simultaneous observation of all these lines alongside the Mg {\sc II} h and k lines 
would facilitate magnetic diagnostics spanning from the solar photosphere to the 
upper chromosphere \citep{AfonsoDelgado2023bApJ,AfonsoDelgado2023cApJ,
AfonsoDelgado2025ApJ}.

\backmatter

\bmhead{Acknowledgements}

H.L. acknowledges the support from the National Key R\&D Program of 
China (2021YFA1600500, 2021YFA1600503), and the National Natural 
Science Foundation of China under grant No. 12473051. 
T.P.A. and J.T.B. acknowledge support from the Agencia Estatal de
Investigaci\'on del Ministerio de Ciencia, Innovación y Universidades
(MCIU/AEI) under grant ``Polarimetric Inference of Magnetic Fields''
and the European Regional Development Fund (ERDF) with reference
PID2022-136563NB-I00/10.13039/501100011033.
T.P.A.'s participation in the publication is part of the Project
RYC2021-034006-I, funded by MICIN/AEI/10.13039/501100011033, and the
European Union ``NextGenerationEU''/RTRP.

\section*{Conflicts of interest}

All authors declare that they have no conflicts of interest.

\bibliography{RvMPP}

\end{document}